\documentclass[10pt,final,letterpaper,twocolumn,romanappendices]{IEEEtran}
\usepackage{cite}
\usepackage{epstopdf}
\usepackage{epsfig}
\usepackage{amsthm}
\usepackage{amsmath,amssymb,amsfonts,amstext,amsbsy,amsopn,dsfont}
\usepackage{cases}
\usepackage{sublabel}
\usepackage{color}
\usepackage{array}

\newtheorem{theorem}{\textbf{Theorem}}
\newtheorem{proposition}{\textbf{Proposition}}

\newtheorem{remark}{\textbf{Remark}}

\newcommand{\defn}{\triangleq}
\newcommand{\dif}{\textmd{d}}
\newcommand{\subB}{\mathsf{B}}
\newcommand{\subU}{\mathsf{U}}
\newcommand{\subC}{\mathsf{C}}
\newcommand{\subON}{\mathrm{on}}
\newcommand{\subOFF}{\mathrm{off}}
\newcommand{\subVoid}{\emptyset}

\begin{document}

\title{Optimal Cell Load and Throughput in Green Small Cell Networks with Generalized Cell Association}

\author{Chun-Hung Liu and Li-Chun Wang
\thanks{C.-H. Liu and L.-C. Wang are with the Department of Electrical and Computer Engineering at National Chiao Tung University, Hsinchu, Taiwan. The contact author is Dr. Liu  (e-mail: chungliu@nctu.edu.tw). Manuscript last modified: \today.}
}

\maketitle

\begin{abstract}
This paper thoroughly explores the fundamental interactions between cell association, cell load and throughput in a green (energy-efficient) small cell network in which all base stations form a homogeneous Poisson point process (PPP) of intensity $\lambda_{\subB}$ and all users form another independent PPP of intensity $\lambda_{\subU}$. Cell voidness, usually disregarded due to rarity in cellular network modeling, is first theoretically analyzed under generalized (channel-aware) cell association (GCA). We show that the void cell probability cannot be neglected any more since it is bounded above by $\exp(-\lambda_{\subU}/\lambda_{\subB})$ that is typically not small in a small cell network. The accurate expression of the void cell probability for GCA is characterized and it is used to derive the average cell and user throughputs. We learn that cell association and cell load $\lambda_{\subU}/\lambda_{\subB}$ significantly affect these two throughputs.  According to the average cell and user throughputs, the green cell and user throughputs are defined respectively to reflect whether the energy of a base station is efficiently used to transmit information or not. In order to achieve satisfactory throughput with certain level of greenness, cell load should be properly determined. We present the theoretical solutions of the optimal cell loads that maximize the green cell and user throughputs, respectively, and verify their correctness by  simulation.   
\end{abstract}


\section{Introduction}
 
In recent years, we have witnessed a trend that mobile devices, such as smart phones and tablets, have been proliferating and relentlessly penetrating our daily life. Such powerful handsets have created a new dimension of transmitting information over newly developed wireless technologies. Accordingly, this situation has urged customers to expect more and more throughput to suffice their versatile wireless demands. To boost network throughput, one of the most effective approaches is to increase the spatial frequency reuse by deploying more cells in a given area so that there are fewer users sharing time and/or frequency resources in a cell \cite{NBJLDM14}. Furthermore, as cell intensity (density) increases, users are able to connect to a closer base station and thus desired signals suffer less path loss. Therefore, a small cell network consisting of highly dense picocells and/or femtocells has gained much attention since small cell base stations (BSs) can be deployed indefinitely and the achievable throughput seems not to have a hard limit.

Migrating the topology of traditional macro cells to the one of small cells is the trend for the next generation of cellular systems since a small cell BS has much less power consumption than a traditional macro cell BS whose power consumption typically accounts for 80\% of the total power consumption of a cellular system \cite{IHBSSSS13}. Cellular networks with a small cell topology absolutely help the entire information and communication technology (ICT) industry lessen CO$_2$ emission, which is estimated to account for about 6\% of the global emission in 2020 \cite{AFGFJMGB11}. Small cell BSs also have an advantage of low operational cost and hardware complexity. For example, they can quickly switch their different power control modes for saving power\cite{IAFBLH11}. Due to dense deployment, however, small cell networks seriously suffer the \textit{void cell} issue that is hardly noticed in the traditional marco BS network. The void cell problem originates user-centric cell association\footnote{User-centric cell association means that every user in the network tries to associate with its best service BS via some cell association (or called user association in some literatures) schemes. For example, if all users are looking for a BS that is able to provide the long-term strongest signal power, they will associate with their nearest BS if fading and/or shadowing effects are completely removed from their channels.} and it could give rise to no users in a cell. To explain this cell voidness phenomenon, consider there are $m\times n$ users uniformly distributed in a unit-area network which is tessellated by $n$ equal-sized cells of BSs and thus the user intensity is $m\times n$ and the BS intesnity is $n$.  The probability of having no users in a cell is $\left(1-\frac{1}{n}\right)^{mn}$ which is close to $e^{-m}$ as $n$ is very large. This tells us that the void cell probability is not negligible any more  in a small cell network because such a network usually has a small $m$, i.e., the ratio of the user intensity to the BS intensity is not large. 

\subsection{Motivation and Prior Work}
Without modeling the impact of void cells, the analyses of network performance metrics, e.g., coverage probability and average throughput, are apparently underestimated since the void BSs do not contribute any interference. Despite the important phenomenon of void cells, there are very few works that study and model the effect of void cells. Almost all the prior works on cell association in cellular networks overlook this problem \cite{JGAFBRKG11,HSDRKGFBJGA12,HSJYJSXPJGA12,PXCHLJGA13,MDRAGGEC13,SSJGA14}. References \cite{SLKH12,CLJZKBL14,CTPCHLLCW15} do consider the void cell impact on their models, however, the void cell probability they used is only valid for nearest BS association. Although more recent works in \cite{HSMSMWTQSQ14} and \cite{HSMWMSTQSQ15} also incorporate the void cell issue in their models, their cell void probabilities are derived based on the cell association scheme with constant biased weights and they are no longer valid for the cell association schemes with random biased weights.    

Since cell voidness essentially cannot be completely avoided and small cell BSs can easily switch their power control modes, the void BSs can be put in a ``dormant'' mode for saving power.  Such energy-saving strategy entails another question, that is, how much denseness of BS deployment a small cell network needs to optimally exploit the performance of the energy-saving strategy. A very much dense small cell network is absolutely not green (energy-efficient) because its per-unit power throughput is significantly weakened due to too much interference. Previous works on designing the BS intensity with greenness in a heterogeneous or small cell network are still fairly minimal. The work in \cite{TQSQWCCMK11} numerically showed that there exists an optimal ratio of picocell to macro BS intensity that maximizes  energy-efficient throughput without giving a rigorous theoretical explanation. In \cite{DCSZZN13}, a closed-form upper bound on the optimal BS intensity subject to the constraints on the user outage rate was found without considering the energy consumption of a BS. 

The two aforementioned works did not study the optimal BS intensity from the viewpoint of an achievable green throughput. Although the average achievable throughput of a link was investigated in \cite{JGAFBRKG11} and \cite{MDAGGEC13}, it fails to characterize how energy is efficiently used to transported information on the per cell or per user basis. References \cite{HSJYJSXPJGA12} and \cite{SRCWC13} also present some approximated average per-user achievable throughputs in a cell that do not consider the void cell issue and are only valid for nearest BS association with constant biased weights. In \cite{YSSTQSQMKHS13}, an energy-efficient throughput metric was proposed based on the concept of outage capacity. None of these prior works on  the average per-cell or per-user throughputs investigate the optimal cell load problem as well as consider the void cell issue in their models.

\subsection{Contributions}
 To characterize the void cell probability in a small cell network with spatial randomness,  in this paper we consider a single-tier small cell network in which all BSs form a homogeneous Poisson point process (PPP) of intensity $\lambda_{\subB}$ and all users form another independent PPP of intensity $\lambda_{\subU}$. The fundamental interaction between user-centric cell association and void cell probability is first delved. All users associate with their serving BS by the proposed generalized cell association (GCA) scheme that is able to cover several cell association schemes, such as nearest cell association, maximum received power association, etc. We theoretically show that the achievable lower bound on the void probability of a cell is $\exp\left(-\frac{\lambda_{\subU}}{\lambda_{\subB}}\right)$, where $\lambda_{\subU}/\lambda_{\subB}$ is termed ``cell load''. An accurate closed-form expression for the void cell probability of GCA is derived under a more practical channel model that incorporates the effects of path loss, Rayleigh fading and log-normal shadowing. The derived void cell probability provides us some insights on how the void cell probability is affected by channel impairments and how its theoretical lower bound can be achieved by GCA. To the best of our knowledge, the fundamental relationship between random cell association and void cell probability is first studied in this work.

Under the assumption that  all the associated BSs form a thinning homogeneous PPP, the average cell and user throughputs are proposed to capture the impact of cell voidness. The average cell throughout is a per-cell average throughput metric, whereas the average user throughput is a per-user average throughput metric in a cell. The near closed-form expressions for these two throughputs parameterized by cell load $\lambda_{\subU}/\lambda_{\subB}$ are derived and they are the neatest results than any other similar throughput results in the literature. Most importantly, they directly reflect that the GCA scheme favoring large channel power can significantly benefit them. Also, we show that there exists a unique optimal cell load that maximizes the average user throughput and it is accurately equal to the fixed point of a special function pertaining to the void cell probability. 

A simple green power control, which switches a BS between active and dormant modes depending on whether the BS is void or not, is used to help the network save energy. The green cell and user throughputs are proposed and they aim to characterize how green the per-cell and per-user throughputs can be in the network. We show that both the green cell and user throughputs can be maximized by a sole optimal cell load that can be accurately acquired by calculating the fixed point of the derived special function. Once the optimal cell load is found, the optimal BS intensity corresponding to any given user intensity is also readily obtained and deploying BSs based on this intensity can attain the maximum energy efficiency of information transportation. From these derived results, we can conclude that green power control schemes that keep large power consumption difference between active and dormant modes and cell association schemes that favor high channel power both significantly improve the optimal cell load and green throughputs. 

\subsection{Paper Organization}
The rest of this paper is organized as follows. In Section \ref{Sec:SystemModel}, the system model of the small cell network with generalized cell association is introduced and the preliminary results of the random conservation property of a general PPP and the power consumption model of a small cell base station are presented. Section \ref{Sec:CellVoidProb} studies the void probability of a cell, cell load, average cell and user throughputs. The results of optimal green cell load and green cell and user throughputs are elaborated in Section \ref{Sec:OptGreenIntesityThrou}. Finally, Section \ref{Sec:Conclusion} concludes our important findings and observations in this paper.

\section{System Model and Preliminaries}\label{Sec:SystemModel}
\subsection{Network and Cell Association Models}\label{Sec:NetCellAssModel}
We consider an infinitely large and planar small cell network in which all users form a homogeneous Poisson point process (PPP) $\Phi_{\subU}$ of intensity $\lambda_{\subU}$ whereas one-tier  (small-sized) base stations that provide service to all users independently form another \textit{marked} homogeneous PPP $\Phi_{\subB}$ of intensity $\lambda_{\subB}$ given by
\begin{align}\label{Eqn:PPPModelBS}
\Phi_{\subB}\defn &\{(B_i, H_i, \mathcal{C}_i, V_i): B_i\in\mathbb{R}^2,  H_i\in\mathbb{R}_{++}, \mathcal{C}_i\subset\mathbb{R}^2, \nonumber\\
& V_i\in\{0,1\}, \forall i\in\mathbb{N}_+\},
\end{align} 
where $B_i$ denotes base station $i$ and its location, $H_i$ is used to characterize the downlink fading and shadowing channel power gain from $B_i$ to its service user, the cell region of $B_i$ is represented by $\mathcal{C}_i$, $V_i$ is a void cell index that indicates whether or not $\mathcal{C}_i$ has a user, i.e., whether $\mathcal{C}_i\cap\Phi_{\subU}=\emptyset$ is true or not -- $V_i$ is equal to one if $\mathcal{C}_i\cap\Phi_{\subU}\neq\emptyset$, otherwise zero. In order to capture stochastic behavior of the channel power in the downlink, we assume that all channel power gains $H_i$'s are i.i.d. random variables and their probability intensity function (pdf) that characterizes the composite effect of Rayleigh fading and log-normal shadowing is given by \cite{STUBER01}
\begin{equation}
f_H(h) = \frac{1}{\sqrt{2\pi \sigma^2_s}} \int_{0^+}^{\infty}\frac{1}{x^2} \exp\left(-\frac{h}{x}-\frac{(\ln x-\mu_s)^2}{2\sigma^2_s}\right) \dif x,  
\end{equation} 
where $\mu_s$ and $\sigma^2_s$ are the mean and variance of log-normal shadowing, respectively.  

\begin{figure}[!t]
  	\centering
  	\includegraphics[width=3.7in,height=2.5in]{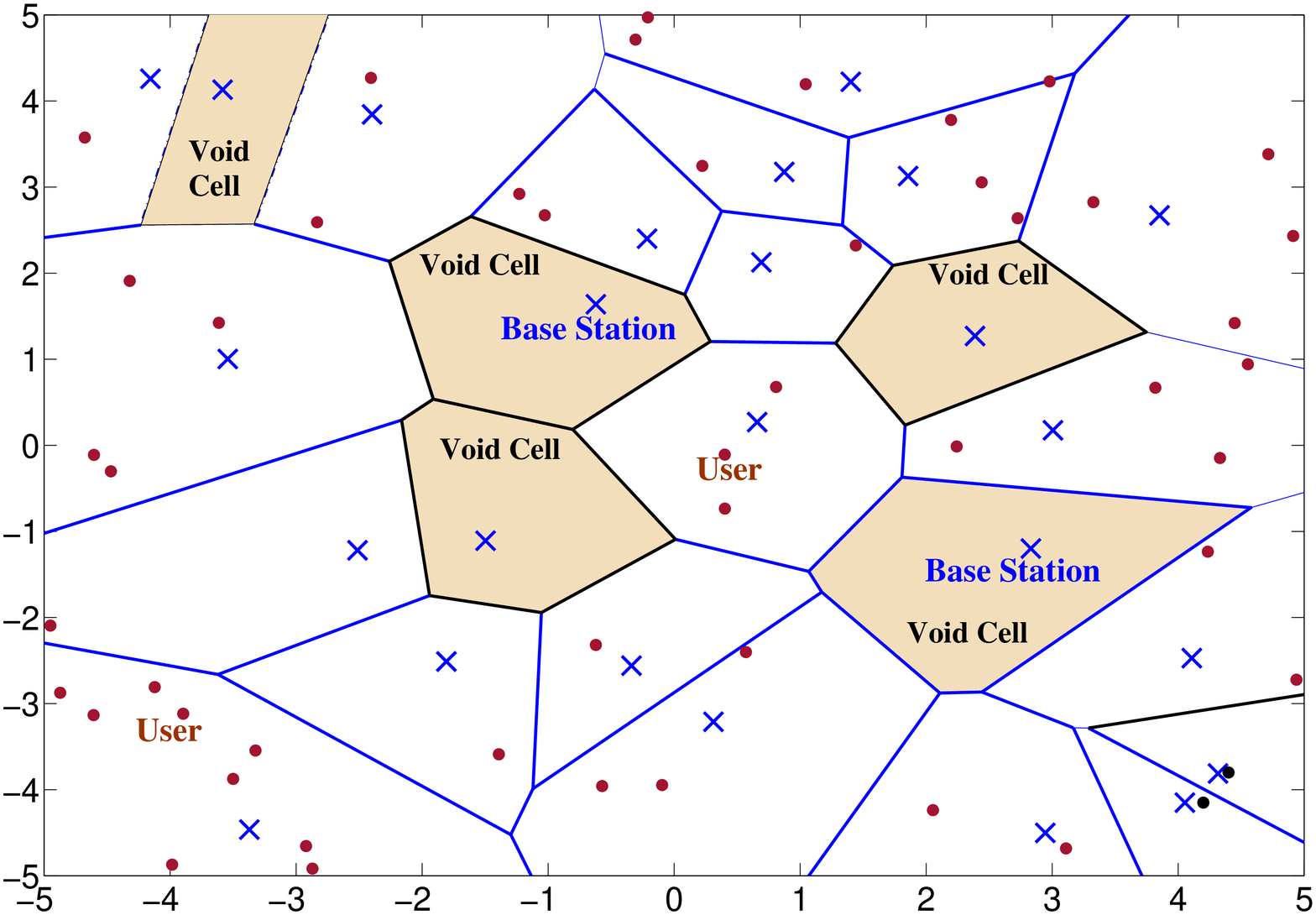}
  	\caption{An illustration example of void cells in a small cell network. BSs (blue crosses) and users (brown dots) are two independent PPPs and the cells are created by Voronoi tessellation. The intensity ratio of users to base stations is $\frac{\lambda_{\subU}}{\lambda_{\subB}}=2$.}
  	\label{Fig:VoidCell}
\end{figure}

Without loss of generality, our following analysis will be based on a typical user $U_0$ located at the origin. Each user associates with a base station in $\Phi_{\subB}$ by using a  generalized (channel-aware) cell association (GCA) scheme, i.e.,  user $U_0$ associates with its serving base station $B^*_0$ via the following scheme
\begin{equation}\label{Eqn:GCA}
B_0^*=\arg\sup_{B_i\in\Phi_{\subB}} \left(W_iH_i\|B_i\|^{-\alpha}\right),
\end{equation}
where all $W_i$'s are the i.i.d.  random association weights for BSs, $\alpha>2$ is the path loss exponent and $\|B_i\|$ denotes the Euclidean distance between $B_i$ and the origin. The motivation of proposing the GCA scheme in \eqref{Eqn:GCA} is two-fold. First, GCA can generally cover cell association schemes with deterministic and/or random association weights. For example, if the channel power gain $H_i$ is available and $W_i= 1/H_i$, GCA reduces to \textit{nearest} BS association. GCA becomes \textit{maximum received power} association provided that all $W_i$'s are the same constant. Second, GCA can be viewed as an adaptable cell association scheme and it is suitable no matter whether BSs can acquire the mean channel power gains of users in time or not. For example, BSs may not be able to estimate the mean received power from non-stationary users that are moving very fast \cite{STUBER01}. GCA is essentially a ``user-centric'' scheme, that is, it is able to ensure every user to connect to a certain base station. No users are blocked out of the network. Nonetheless, user-centric cell association cannot, as we will show later, guarantee every cell is associated with at least one user, i.e., the probability of a void cell (a cell without users) is always bounded above zero and non-negligible, especially in a small cell network with a large BS intensity. A simulation example of the void cell phenomenon in a small cell network with Voronoi tessellation is illustrated in Fig. \ref{Fig:VoidCell}.

\subsection{Random Conservation Property of a General PPP}
In this subsection, we introduce the random conservation property of a general PPP, which specifies how the intensity measure of a PPP is changed after all points of the PPP are transformed by i.i.d. random mapping matrices. This property is elaborated in the following theorem.
\begin{theorem}[Random Conservation Property of a General  PPP]\label{Thm:RanTransProp}
Suppose $\Phi'$ is a marked general PPP of intensity measure $\Lambda'$ on $\mathbb{R}^{\mathrm{d}'}$, which can be expressed as follows
\begin{equation}
\Phi'\defn\{(X'_i, \mathbf{T}_i): X'_i\in\mathbb{R}^{\mathrm{d}'}, \mathbf{T}_i\in\mathbb{R}^{\mathrm{d}\times \mathrm{d}'}, \forall i\in\mathbb{N}_{+}\},
\end{equation}
where $X'_i$ denotes node $i$ and its location and $\mathbf{T}_i: \mathbb{R}^{\mathrm{d}'}\rightarrow\mathbb{R}^{d}$ is the non-singular mapping matrix (operator) of node $X'_i$. For all $i\neq j$, $\mathbf{T}_i$ and $\mathbf{T}_j$ are two different random matrices and their corresponding elements are i.i.d. random variables. Let $\hat{\Phi}$ be the mapped point process on $\mathbb{R}^{\mathrm{d}}$ generated by using the random mapping matrix of each node in $\Phi'$, i.e., it is defined as
\begin{equation}
\hat{\Phi}\defn\{\hat{X}_i\defn\mathbf{T}_i (X'_i): X'_i\in\Phi', \mathbf{T}_i\in\mathbb{R}^{\mathrm{d}\times \mathrm{d}'}, \forall i\in\mathbb{N}_{+}\}.
\end{equation}
For any $\mathrm{d}$-dimensional bounded Borel set $\mathcal{A}\subset \mathbb{R}^{\mathrm{d}}$  and $\mathrm{d}'$-dimensional bounded Borel set $\mathcal{A}'\subset \mathbb{R}^{\mathrm{d}'}$, if $\nu_{\mathrm{d}'}(\mathcal{A}')=\nu_{\mathrm{d}}(\mathcal{A})$, then $\hat{\Phi}$ is a general PPP of intensity measure $\hat{\Lambda}$ given by
\begin{equation}\label{Eqn:TransGenIntensityMeasure}
\hat{\Lambda}(\mathcal{A}) = \Lambda'(\mathcal{A}')\mathbb{E}\left[\frac{1 }{\sqrt{\det\left(\mathbf{T}^{\mathrm{T}}\mathbf{T}\right)}}\right],
\end{equation}
where $\mathbf{T}^{\mathrm{T}}$ is the transpose of $\mathbf{T}$. If $\Phi'$ is homogeneous with intensity $\lambda'$, $\hat{\Phi}$ is also homogeneous and has the following intensity 
\begin{equation}\label{Eqn:TransHomoIntensityMeasure}
\hat{\lambda} = \lambda'\, \mathbb{E}\left[\frac{1}{\sqrt{\det(\mathbf{T}^{\mathrm{T}}\mathbf{T}) }}\right].
\end{equation}
\end{theorem}
\begin{IEEEproof}
See Appendix \ref{App:ProofRanTransProp}.
\end{IEEEproof}
\begin{remark}
Theorem \ref{Thm:RanTransProp} is a generalization of the conservation property in \cite{DSWKJM96}\cite{CHLBRSC15}. In a special case of $\mathrm{d}=\mathrm{d}'=2$, all points in $\hat{\Phi}$ are mapped from their corresponding points in a homogeneous PPP $\Phi'$ by scaling them with i.i.d. diagonal random matrices $\mathbf{T}_i=\text{diag}(T_i, T_i)$. In this case, $\hat{\lambda}$ is equal to $\lambda' \mathbb{E}[T^{-2}]$.  
\end{remark}

The random conservation property can significantly reduce the complexity of analyzing the statistics of some performance metrics induced by a PPP, especially a homogeneous PPP with i.i.d. marks. Hereupon the GCA scheme in \eqref{Eqn:GCA} can be rewritten as $B_0^*=\arg\inf_{B_i\in\Phi_{\subB}} \|(W_iH_i)^{-\frac{1}{\alpha}}B_i\|$ which can be further simplified by Theorem \ref{Thm:RanTransProp} as
\begin{equation}\label{Eqn:EquAssoBSofGCA}
\|B_0^*\|\stackrel{d}{=} (WH)^{\frac{1}{\alpha}}\left(\inf_{\hat{B}_i\in\hat{\Phi}_{\subB}} \|\hat{B}_i\|\right),
\end{equation}
where $\stackrel{d}{=}$ stands for equivalence in distribution,  $\hat{B}_i\stackrel{d}{=} (W_iH_i)^{-\frac{1}{\alpha}}B_i$ and $\hat{\Phi}_{\subB}$ is a homogeneous PPP of intensity $\lambda_{\subB}\mathbb{E}\left[(WH)^{\frac{2}{\alpha}}\right]$ based on \eqref{Eqn:TransHomoIntensityMeasure} in Theorem \ref{Thm:RanTransProp}. Hence, the distribution of the distance between user $U_0$ and its serving BS $B^*_0$ can be instead equivalently found by the distribution of the $(WH)^{\frac{1}{\alpha}}$-weighted distance from the origin to the \textit{nearest} BS in the new PPP $\hat{\Phi}_{\subB}$. In other words, \textit{the random transformation property can transform GCA into another form of nearest BS association}, which significantly simplifies the analysis of the performance metric of GCA, such as coverage/outage probability, since many existing results of nearest BS association can be applied in this context by simply modifying them with an updated intensity of the BSs.  

\subsection{Power Consumption Model of a Small Cell BS}\label{SubSet:PowConModel}
Although the cell voidness issue mentioned in Section \ref{Sec:NetCellAssModel} seems inevitable, its impact on network energy consumption can be alleviated by applying green power control at BSs. The green power control scheme for a small cell BS has two modes -- the active mode is for non-void BSs, whereas the dormant mode is for void BSs. For base station $B_i$, the power consumption of such green power control can be characterized by the following expression \cite{GAVGCD11}
\begin{equation}\label{Eqn:GreenPowControl}
\Psi_i=V_i(P_{\subON}+\delta P_t)+(1-V_i)P_{\subOFF},
\end{equation}
where $\Psi_i$ is the power consumed by BS $B_i$, $P_{\subON}$ is the power consumed by the hardware of a BS in the active mode, $P_t$ denotes the constant transmit power of a base station, $\delta>0$ is a scaling constant for the transmit power which usually depends on the power amplifier used by a BS, and $P_{\subOFF}$ denotes the power consumption of a dormant BS. Also, we assume $P_{\subON}>P_{\subOFF}$ since it is usually the case in practice.  
The average power consumption in \eqref{Eqn:GreenPowControl} can be respectively written as
\begin{equation}\label{Eqn:AvgComPower}
\mathbb{E}[\Psi]  = (1-p_{\emptyset}) (P_{\subON}+\delta P_t) +p_{\subVoid}P_{\subOFF}
\end{equation}
in which $p_{\subVoid}\defn\mathbb{P}[V_i=0]$ represents the void probability of a cell and its analytical result will be elaborated in Section \ref{Sec:CellVoidProb}. Note that $p_{\subVoid}$ is certainly affected by cell association schemes adopted by users. In Section  \ref{Sec:CellVoidProb}, we will investigate how to characterize $p_{\subVoid}$ for GCA in \eqref{Eqn:GCA} from a fundamental connectivity perspective. In addition, the transmit power $P_t$ of all BSs is assumed to be able to provide the minimum received power  $P_{\min}$ to their associated users by compensating the mean path loss, i.e., $P_t=P_{\min}\inf\{\mathbb{E}[\|B^*_0\|^{\alpha}]\}$. According to $\|B_0^*\|$ in \eqref{Eqn:EquAssoBSofGCA}, we have
\begin{align*}
\mathbb{E}[\|B^*_0\|^{\alpha}]=& \mathbb{E}\bigg[2\pi   \lambda_{\subB}\mathbb{E}[(WH)^{\frac{2}{\alpha}}](WH)^{-\frac{2}{\alpha}}\\
&\int_{0}^{\infty} r^{1+\alpha}e^{-\pi\lambda_{\subB}\mathbb{E}[(WH)^{\frac{2}{\alpha}}](WH)^{-\frac{2}{\alpha}} r^2} \dif r\bigg] \nonumber\\
=&\mathbb{E}\left[\frac{ \Gamma(1+\frac{\alpha}{2})}{(\pi\lambda_{\subB}(WH)^{-\frac{2}{\alpha}}\mathbb{E}[(WH)^{\frac{2}{\alpha}}])^{\frac{\alpha}{2}}} \right] \geq\frac{ \Gamma(1+\frac{\alpha}{2})}{(\pi\lambda_{\subB}\zeta)^{\frac{\alpha}{2}}},
\end{align*}
where $\Gamma(x)=\int_{0}^{\infty} t^{x-1}e^{-t}\dif t$ is the Gamma function and $\zeta\defn \mathbb{E} \left[(WH)^{\frac{2}{\alpha}}\right]\mathbb{E} \left[(WH)^{-\frac{2}{\alpha}}\right]\geq 1$. Thus, $P_t$ is given by
\begin{equation}\label{Eqn:ReqTxPower}
P_t=\frac{P_{\min} \Gamma(1+\frac{\alpha}{2})}{(\pi\lambda_{\subB}\zeta)^{\frac{\alpha}{2}}}.
\end{equation}
Note that $P_t$ in (11) is implicitly assumed not to exceed the maximum transmit power provided by a small cell BS for a given $P_{\min}$ and $\lambda_{\subB}$. By plugging \eqref{Eqn:ReqTxPower} into \eqref{Eqn:AvgComPower}, the explicit result of $\mathbb{E}[\Psi] $ is acquired, which will be used to define the green cell and user throughputs in Section \ref{Sec:OptGreenIntesityThrou}.
 
\section{Void Cell Probability, Cell Load and Average Throughputs}\label{Sec:CellVoidProb}
\subsection{The Void Probability of a Cell for GCA}
As pointed out in the previous section, the GCA scheme or any ``user-centric'' association scheme cannot guarantee that there are at least one user in each cell, i.e., a void cell could exist in the network. This phenomenon can be intuitively interpreted by using a Poisson-Dirichlet (Voronoi) tessellation for a PPP of BSs. Suppose the Voronoi tessellation is used to determine the cell of each BS in $\Phi_{\subB}$ and all users adopt the nearest BS association scheme to connect with their serving BS. This nearest associating process can be viewed as the process of dropping all users in $\Phi_{\subU}$ on the large plane consisting of the Voronoi-tessellated cells formed by $\Phi_{\subB}$. Under this circumstance, the probability mass function (pmf) of the number of users in a cell of $\Phi_{\subB}$ can be expressed as 
\begin{equation}\label{Eqn:PmfNumUserVorCell}
p_n\defn\mathbb{P}[\Phi_{\subU}(\mathcal{C})=n] = \mathbb{E}\left[\frac{(\lambda_{\subU}\nu(\mathcal{C}))^n}{n!}e^{-\lambda_{\subU}\nu(\mathcal{C})}\right],
\end{equation}
where $\mathcal{C}$ denotes a Voronoi cell of a BS in $\Phi_{\subB}$, $\Phi_{\subU}(\mathcal{C})$ represents the number of users in cell $\mathcal{C}$, and $\nu(\mathcal{C})$ is the Lebesgue measure of $\mathcal{C}$.

Unfortunately, the theoretical result of $p_n$ in \eqref{Eqn:PmfNumUserVorCell} is unknown since the pdf of a Voronoi cell area is still an open problem \cite{DSWKJM96}. However, it can be accurately approximated by using a Gamma distribution with some particular parameters \cite{DSWKJM96,JSFNZ07}. Reference \cite{JSFNZ07} suggests the following Gamma distribution for $f_{\nu(\mathcal{C})}(x)$:
\begin{equation}\label{Eqn:ApporxPdfVoronoiCell}
f_{\nu(\mathcal{C})}(x) \approx \frac{(\hat{\rho} \lambda_{\subB}x)^{\hat{\rho}}}{\Gamma(\hat{\rho})x}e^{-\hat{\rho}\lambda_{\subB}x} 
\end{equation} 
and $\hat{\rho}=\frac{7}{2}$ can achieve an accurate pdf of a Voronoi cell area. Substituting \eqref{Eqn:ApporxPdfVoronoiCell} into  \eqref{Eqn:PmfNumUserVorCell} yields the following result:   
\begin{align}
p_n
&\approx\frac{\lambda^n_{\subU}}{n!}\frac{(\hat{\rho} \lambda_{\subB})^{\hat{\rho}}}{\Gamma(\hat{\rho})}\int_{0}^{\infty}   x^{n+\hat{\rho}-1}  e^{-(\hat{\rho}\lambda_{\subB}+\lambda_{\subU})x}\dif x\nonumber\\ &=\frac{1}{n!}\frac{\Gamma(n+\hat{\rho})}{\Gamma(\hat{\rho})}\left(\frac{\lambda_{\subU}/\lambda_{\subB}}{\hat{\rho}+\lambda_{\subU}/\lambda_{\subB}}\right)^n\left(1+\frac{\lambda_{\subU}}{\hat{\rho}\lambda_{\subB}}\right)^{-\hat{\rho}}.\label{Eqn:AppPmfNumUserVorCell}
\end{align}
We call the term $\lambda_{\subU}/\lambda_{\subB}$ in \eqref{Eqn:AppPmfNumUserVorCell} the \textit{cell load} of the network since it represents the mean number of users in a Voronori-tessellated cell\footnote{Although the pdf of Voronori cells for a homogeneous PPP of intensity $\lambda$ is unknown, its mean can be shown as $1/\lambda$\cite{DSWKJM96}. Thus, cell load $\defn\mathbb{E}[\lambda_{\subU}\nu(\mathcal{C}_i)]=\lambda_{\subU}/\lambda_{\subB}$.}. Hence, the pmf of the number of users in a cell for nearest BS association has an accurate closed-form expression. The void probability of a cell, $p_{\emptyset}$,  can be found by $p_n$ with  the case of $n=0$, which is the last term of $p_n$ in \eqref{Eqn:AppPmfNumUserVorCell}, i.e.,
\begin{equation}\label{Eqn:VoidProbNearBS}
p_{\emptyset}\defn \mathbb{P}[V_i=0]=\left(1+\frac{\lambda_{\subU}}{\hat{\rho}\lambda_{\subB}}\right)^{-\hat{\rho}}
\end{equation}
and this indicates that the intensity of the void BSs is $\lambda_{\subB}p_{\emptyset}$ that is not negligible especially when cell load $\lambda_{\subU}/\lambda_{\subB}$ is small. Most importantly, \textit{the results in \eqref{Eqn:ApporxPdfVoronoiCell} and \eqref{Eqn:AppPmfNumUserVorCell} are no longer accurate for all non-nearest BS association schemes since users do not necessarily associate with their nearest BS, such as the GCA scheme proposed in Section \ref{Sec:SystemModel}}. However, an accurate void probability of a BS for GCA can be derived as shown in the following. By Jensen's inequality,  the lower bound on $p_{\emptyset}$ in \eqref{Eqn:VoidProbNearBS} is given by
\begin{equation}\label{Eqn:LowBoundVoidProb}
p_{\emptyset}=\mathbb{E}\left[e^{-\lambda_{\subU}\nu(\mathcal{C})}\right] \geq \exp\left(-\frac{\lambda_{\subU}}{\lambda_{\subB}}\right).
\end{equation}
This lower bound on the void probability reveals three crucial implications: (i) the void probability of a cell is always bounded above zero  such that there could be a certain number of void BSs in the small cell network; (ii) nearest BS association cannot achieve this lower bound since its void probability in \eqref{Eqn:VoidProbNearBS} is always larger than the lower bound; (iii) from an energy-saving perspective, the lower bound represents the minimum percentage of void BSs in the network that can be turned off to save energy. Later, we will theoretically show that, \textit{this lower bound can be achieved by GCA}. To derive an accurate void probability of a cell for the GCA scheme, we approach this problem from a fundamental connectivity point of view and derive the bounds on $p_{\emptyset}$ as shown in the following theorem.
 
\begin{theorem}\label{Thm:BoundVoidProb}
If all users adopt the GCA scheme defined in \eqref{Eqn:GCA} to associate with a BS in $\Phi_{\subB}$ defined in \eqref{Eqn:PPPModelBS},  the bounds on the void probability of a cell are given by
\begin{equation}\label{Eqn:BoundVoidProb}
\left(1+\frac{\lambda_{\subU}}{\lambda_{\subB}\zeta}\right)^{-\zeta}\geq p_{\emptyset}\geq \exp\left(-\frac{\lambda_{\subU}}{\lambda_{\subB}}\right)
\end{equation} 
and  $\zeta=\mathbb{E}\left[(WH)^{\frac{2}{\alpha}}\right]\mathbb{E}\left[(WH)^{-\frac{2}{\alpha}}\right]\geq 1$ as defined in Section \ref{SubSet:PowConModel}.
 \end{theorem} 
 \begin{IEEEproof}
 See Appendix \ref{App:ProofBoundVoidProb}.
 \end{IEEEproof}
 
According to the proof of Theorem \ref{Thm:BoundVoidProb}, the lower bound on $p_{\emptyset}$ is derived while considering the completely independence exists between the non-associated events of a BS whereas the upper bound is obtained by approaching the opposite case, i.e., all non-associated events of a BS are caused by considering all users in a cell located the same farthest distance from the BS  and thus they are highly correlated. Accordingly, it is reasonably to conjecture that the upper bound is tightly close to the lower bound provided that the cross-correlations between all non-associated events are significantly weakened.  On the other hand, mathematically we know $\lim_{\zeta\rightarrow\infty} \left(1+\frac{\lambda_{\subU}}{\lambda_{\subB}\zeta}\right)^{-\zeta}=\exp\left(-\frac{\lambda_{\subU}}{\lambda_{\subB}}\right)$ and thus the bounds in \eqref{Eqn:BoundVoidProb} are fairly tight as $\zeta$ becomes large. This intuitively reveals that large $\zeta$ will  ``decorrelate'' all non-associated events, which is an important observation since it lets us realize that the lower bound on $p_{\emptyset}$ is easily achieved by enlarging the $\frac{2}{\alpha}$-fractional moment of $W_iH_i$. For example, the lower bound on $p_{\emptyset}$ can be achieved by the maximum received power association scheme provided that channels have large shadowing power. In addition, the bounds in \eqref{Eqn:BoundVoidProb} can be tight for the case of a large cell load. This indicates that the void probability of a BS is reduced when more users join the network under a given BS intensity. In other words, when the network has a large user population the efficacy of reducing $p_{\emptyset}$ by using large $\zeta$ is apparently undermined such that the performance of GCA is similar to that of nearest BS association in this case. 
  
Although the bounds on $p_{\emptyset}$ are characterized, an accurate result of $p_{\emptyset}$ is still needed since it will help us understand how many BSs per unit area are void and they should be switched to the dormant mode. The following proposition renders an accurate heuristic result of $p_{\emptyset}$ for GCA.  
 \begin{proposition}\label{Prop:VoidCellProbGCA}
 The void probability of a BS can be accurately approximated by
 \begin{equation}\label{Eqn:VoidProbGCA}
 p_{\emptyset} = \left(1+\frac{\lambda_{\subU}}{\rho\lambda_{\subB}}\right)^{-\rho}
 \end{equation}
where $\rho=\frac{7}{2}\mathbb{E}[(WH)^{\frac{2}{\alpha}}]\mathbb{E}[(WH)^{-\frac{2}{\alpha}}]=\frac{7}{2}\zeta$ if the GCA scheme is used in the network. 
 \end{proposition}
 \begin{IEEEproof}
 Since $\left(1+x\right)^{-1}\geq\left(1+x/a\right)^{-a}$ for $a>1$, letting $x = \lambda_{\subU}/ \lambda_{\subB}$ and $a = \rho / \zeta$ leads to
 $$\left(1+\frac{\lambda_{\subU}}{\lambda_{\subB}\zeta}\right)^{-\zeta}>\left(1+\frac{\lambda_{\subU}}{\lambda_{\subB}\rho}\right)^{-\rho}$$
 since $\rho>\zeta\geq 1$. Based on the accurate approximation of the pdf of a Voronoi cell area suggested in \cite{JSFNZ07}, $p_{\emptyset}$ with $\rho=\frac{7}{2}$ in \eqref{Eqn:VoidProbNearBS} is an accurate void probability of a BS for nearest BS association. Therefore, we can conclude that $p_{\emptyset}$ with $\rho=\frac{7}{2}\zeta$ is accurately the void probability of a BS for the GCA scheme since such $\rho$ reduces to $\frac{7}{2}$ as GCA reduces to nearest BS association (i.e., $W_i=1/H_i$ for all $i$).
 \end{IEEEproof}
 
  \begin{figure}[!t]
  	\centering
  	\includegraphics[width=3.65in,height=2.6in]{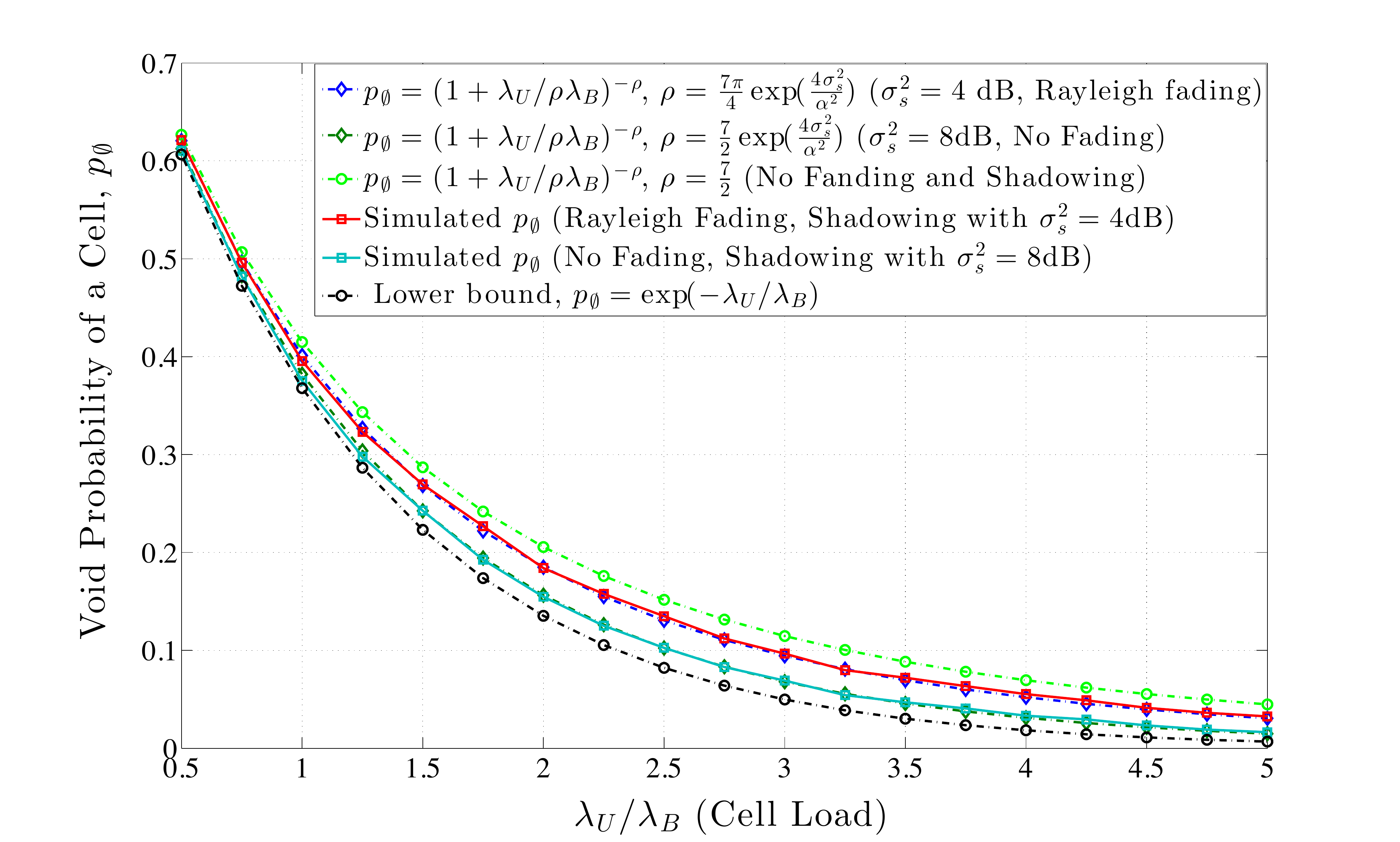}
  	\caption{The void probabilities of a cell for nearest BS association ($W_i=1/H_i$, $\rho=\frac{7}{2}$) and maximum received power association ($W_i=1$, $\rho>\frac{7}{2}$). The network parameters for simulation are path loss exponent $\alpha=3.76$, shadowing mean $\mu_s=0$ dB, $\lambda_{\subU}=370$ users/km$^2$.}
  	\label{Fig:VoidProbCell}
  \end{figure}
 
 Fig. \ref{Fig:VoidProbCell} shows the simulation result of the void probabilities for  nearest BS association and maximum received power association. For nearest BS association, we have $\rho=\frac{7}{2}$ since $W_i=1/H_i$, while $\rho$ for maximum received power association is given by
 \begin{equation}\label{Eqn:RhoMaxRxPowAss}
 \rho= \frac{7}{2}\mathbb{E}\left[H^{\frac{2}{\alpha}}\right]\mathbb{E}\left[H^{-\frac{2}{\alpha}}\right]=\frac{7\pi}{\alpha\sin(2\pi/\alpha)}\exp\left(\frac{4\sigma^2_s}{\alpha^2}\right),
 \end{equation} 
 which only depends on the shadowing power (variance) and path loss exponent. First, we see that the void probability of nearest BS association is no longer accurate if maximum received power association is used.  For example, when $\frac{\lambda_{\subU}}{\lambda_{\subB}}\approx 2$, the void probabilities for the nearest BS and maximum received power association schemes with $8$-dB shadowing are $0.2$ and $0.14$, respectively. The lower bound on the void probability is around 0.135. The void cell probability given in \eqref{Eqn:VoidProbGCA} indeed accurately coincides with the simulated result and is much closer to the lower bound. Next, as expected, large shadowing power (i.e., $\rho\gg 1$) indeed makes the void cell probability approach to its lower bound. Thus, GCA is able to achieve this lower bound. 

\subsection{Average Cell Throughput}
Since void cells could exist in the network and they do not contribute network interference, the throughputs obtained in the previous works overlooking the void cell problem are certainly underestimated if interference is treated as noise. To effectively quantify how the void probability impacts the downlink throughput of a BS, we define the following \textit{average cell throughput} assuming the channel input is Gaussian and interference is treated as noise:
\begin{equation}\label{Eqn:DefnAveCellThroughput}
\mathcal{T}_{\subC} \defn\sup_{B^*_0\in\Phi_{\subB}}\left\{ \mathbb{E}\left[\log_2\left(1+\frac{H_0\|B^*_0\|^{-\alpha}}{I_0}\right)\bigg/\nu(\mathcal{C}^*_0)\right]\right\},
\end{equation}
where $I_0\defn \sum_{B_i\in\Phi_{\subB}\setminus B^*_0}V_iH_i\|B_i\|^{-\alpha}$ is the interference at user $U_0$ assuming all spectrum is fully reused and $\nu(C_0^*)$ is the Lebesgue measure of cell $\mathcal{C}_0^*$ associated by user $U_0$. Average cell throughput   $\mathcal{T}_{\subC}$ characterizes  the trade-off relationship between average downlink channel rate and cell size due to the void probability -- large void probability benefits average channel rate owing to interference reduction, yet it increases the average cell size of the associated BSs that cover the whole network such that more path loss is induced.  The explicit result of $\mathcal{T}_{\subC}$ is given in the following proposition. 
\begin{proposition}\label{Prop:AvgCellThroughput}
If every user associates with its cell by GCA and all the associated BSs are modeled as a homogeneous PPP and let $\mathcal{L}_{Z}(s)\defn\mathbb{E}[e^{-sZ}]$ be the Laplace functional of random variable $Z$, then 
\begin{align}\label{Eqn:LapInter}
\mathcal{L}_{\|B_0^*\|^{\alpha}I_0}(s)=\mathbb{E}\left[\frac{G^2}{G^2+(1-p_{\emptyset})\ell(s,G^2)}\right],
\end{align}
where $p_{\emptyset}$ is given in \eqref{Eqn:VoidProbGCA}, $G^2\defn (WH)^{-\frac{2}{\alpha}}\mathbb{E}[(WH)^{\frac{2}{\alpha}}]$, 
\begin{align*}
\ell(s,\phi) \defn & s^{\frac{2}{\alpha}}\bigg\{\mathbb{E}\left[H^{\frac{2}{\alpha}}\right]\Gamma\left(1-\frac{2}{\alpha}\right)- \\
&\hspace{0.25in}\int_{0}^{\phi s^{-\frac{2}{\alpha}}}\left[1-\mathcal{L}_{H}\left(t^{-\frac{\alpha}{2}}\right)\right]\dif t\bigg\}
\end{align*} 
and $\mathcal{L}_{\|B^*_0\|I_0}(s)\leq  \frac{1}{1+(1-p_{\emptyset})\ell(s,\zeta)/\zeta}$ where $\zeta=\mathbb{E}[G^2]$. For a sufficiently large positive integer $n$, the average cell throughput in \eqref{Eqn:DefnAveCellThroughput} can be accurately found by
\begin{align}\label{Eqn:AvgCellThroughput}
\mathcal{T}_{\subC} =&\frac{\lambda_{\subB}(1-p_{\emptyset})}{(\ln 2)(1-1/\rho)}\times \nonumber\\
&\sum_{i=1}^{n} \int_{0}^{\infty}\frac{\omega_i \dif s}{[s+e^{-(\sqrt{2}\sigma_s x_i+\mu_s)}][1+(1-p_{\emptyset})\ell(s,\zeta)/\zeta]}, 
\end{align} 
where $\omega_i=\frac{2^{n-1}n!}{n^2[\mathsf{H}_{n-1}(x_i)]^2}$, $x_i$'s are the roots of the physicists' version of the Hermite polynomial $\mathsf{H}_n(x)$ \cite{MAIAS72}. If there is no shadowing in the channels, \eqref{Eqn:AvgCellThroughput} reduces to 
\begin{equation}\label{Eqn:AvgCellThroughputNoShadow}
\mathcal{T}_{\subC}=\frac{\lambda_{\subB}(1-p_{\emptyset})}{(\ln 2)(1-1/\rho)} \int_{0}^{\infty} \frac{ \dif s}{(s+1)[1+(1-p_{\emptyset})\ell(s,\zeta)/\zeta]}
\end{equation}
and 
$\ell(s,\zeta) = s^{\frac{2}{\alpha}}\left(\frac{2\pi}{\alpha\sin(2\pi/\alpha)}-\int_{0}^{\zeta s^{-\frac{2}{\alpha}}}\frac{dt}{1+t^{\frac{\alpha}{2}}}\right)$.
\end{proposition}

\begin{IEEEproof}
See Appendix \ref{App:ProofAvgCellThroughput}.
\end{IEEEproof}
\begin{remark}
In order to make the derivations of $\mathcal{T}_{\subC}$ tractable, we assume the associated BSs still form a homogeneous PPP in this paper even though in theory they are no longer a PPP  due to the correlation between them (see the proof of Theorem \ref{Thm:BoundVoidProb}). Nonetheless, our previous work in \cite{CHLLCW15} showed that the associated BS can be still accurately approximated by a homogeneous PPP as long as the cell load is not pretty small.
\end{remark}
\begin{remark}\label{Rem:CovProb}
If the coverage probability is defined as the probability that the SIR of the user is no less than a threshold $s\in\mathbb{R}_{++}$, it follows that  $\mathcal{L}_{\|B_0^*\|^{\alpha}I_0}(s)$ in \eqref{Eqn:LapInter} is the user coverage probability in a cell. The result in \eqref{Eqn:LapInter} is much neat and more general if compared with the coverage probability in prior works since it indicates how void cell probability, fading and shadowing affect user's coverage.
\end{remark}

The average cell throughput in \eqref{Eqn:AvgCellThroughput} is the nearest closed-form result than any other similar average throughput results in literature (such as \cite{JGAFBRKG11}\cite{MDRAGGEC13}). Its physical meaning is the per-cell spectrum efficiency of quantifying how the average spectrum efficiency of associated BSs changes along their average cellular size $1/\lambda_{\subB}(1-p_{\emptyset})$ under the GCA scheme. Especially, it is the first cell-wise throughput result including the impact of the void cell probability. It possesses  several important meanings that are addressed in the following, respectively.
\begin{itemize}
\item \textbf{BS intensity dominates $\mathcal{T}_{\subC}$ in heavy cell load}: For the case of heavy cell load (i.e., $\lambda_{\subU}/\lambda_{\subB}\gg 1$), we have  $p_{\emptyset}\approx 0$. Intuitively, decreasing $p_{\emptyset}$ increases more interference, however, it also induces less path loss in desired signals as well as the average cell size of the associated BSs. In this context, $\mathcal{T}_{\subC}$  is approximated by 
\begin{align}
\hspace{-0.2in}\mathcal{T}_{\subC}\approx &\frac{\lambda_{\subB}}{(\ln 2)(1-1/\rho)}\times  \nonumber\\ &\sum_{i=1}^{n}\int_{0}^{\infty} \frac{\omega_i\dif s}{(s+e^{-(\sqrt{2}\sigma_sx_i+\mu_s)})(1+\ell(s,\zeta)/\zeta)},
\end{align}
which indicates $\mathcal{T}_{\subC}$ becomes a linearly increasing function of $\lambda_{\subB}$ such that the asymptotic scaling law $\mathcal{T}_{\subC}\in\Theta(\lambda_{\subB})$ holds. Thus, under the heavy cell load situation deploying more BSs can directly increase $\mathcal{T}_{\subC}$.   

\item \textbf{User intensity dominates $\mathcal{T}_{\subC}$ in light cell load}: For the case of light cell load (i.e., $\lambda_{\subU}/\lambda_{\subB}\ll 1$), we have  $p_{\emptyset}\approx 1$ and $\mathcal{T}_{\subC}$ is approximated by
\begin{align}
\mathcal{T}_{\subC}\approx &\frac{\lambda_{\subU}}{(\ln 2)(1-1/\rho)}\times \nonumber\\
&\sum_{i=1}^{n}\int_{0}^{\infty}\frac{\lambda_{\subB}\omega_i\dif s}{(s+e^{-(\sqrt{2}\sigma_sx_i+\mu_s)})(\lambda_{\subB}+\lambda_{\subU}\ell(s,\zeta)/\zeta)}.
\end{align}
For a given $\lambda_{\subU}$, we can observe that $\mathcal{T}_{\subC}$ is a monotonic increasing and concave function of $\lambda_{\subB}$ and thus $\mathcal{T}_{\subC}\in\Theta(\lambda_{\subU})$ as $\lambda_{\subB}\rightarrow\infty$. For a given $\lambda_{\subB}$ and $\lambda_{\subU}\rightarrow 0$, we also have $\mathcal{T}_{\subC}\in\Theta(\lambda_{\subU})$. Accordingly, making more users join the network can efficiently boost $\mathcal{T}_{\subC}$ in the light cell load situation. This scaling law corrects the flawed scaling result of the average throughput in the previous work without considering void probability\footnote{Using the average rate without considering void probability in \cite{JGAFBRKG11} to calculate $\mathcal{T}_{\subC}$ defined here, only the scaling law $\mathcal{T}_{\subC}\in\Theta(\lambda_{\subB})$ for all $\lambda_{\subU}\in\mathbb{R}_{++}$ can be concluded. This is certainly not true for light cell load since a large portion of BSs are not connected in this case so that they do not contribute any throughput in the network.}. 

\item \textbf{Cell association significantly impacts $\mathcal{T}_{\subC}$}: Cell association schemes have a considerable impact on $\mathcal{T}_{\subC}$ since $p_{\emptyset}$ and $\ell(s,\zeta)$ in \eqref{Eqn:LapInter} both  are affected by cell association weights. Specifically, a cell association scheme with fairly large $\zeta$ certainly benefits $\mathcal{T}_{\subC}$ since $\ell(s,\zeta)/\zeta$ decreases considerably and $(1-p_{\emptyset})$ almost remains a constant in this case. General speaking, nearest cell association ($W_iH_i=1$ and $\zeta=1$) is unable to achieve higher $\mathcal{T}_{\subC}$ than any other cell association schemes (with random  $W_iH_i$ and $\zeta\gg1$). In maximum received power association, for example, cell association results are highly impacted by $\zeta$ that exponentially increases with shadowing power as shown in \eqref{Eqn:RhoMaxRxPowAss} such that $\mathcal{T}_{\subC}$ becomes much larger in a higher shadowing power environment. In this context,  cell association schemes that favor  BSs with stronger channel power surely benefits $\mathcal{T}_{\subC}$. 
\end{itemize}

\begin{figure}[t!]
\centering
\includegraphics[width=3.65in,height=2.5in]{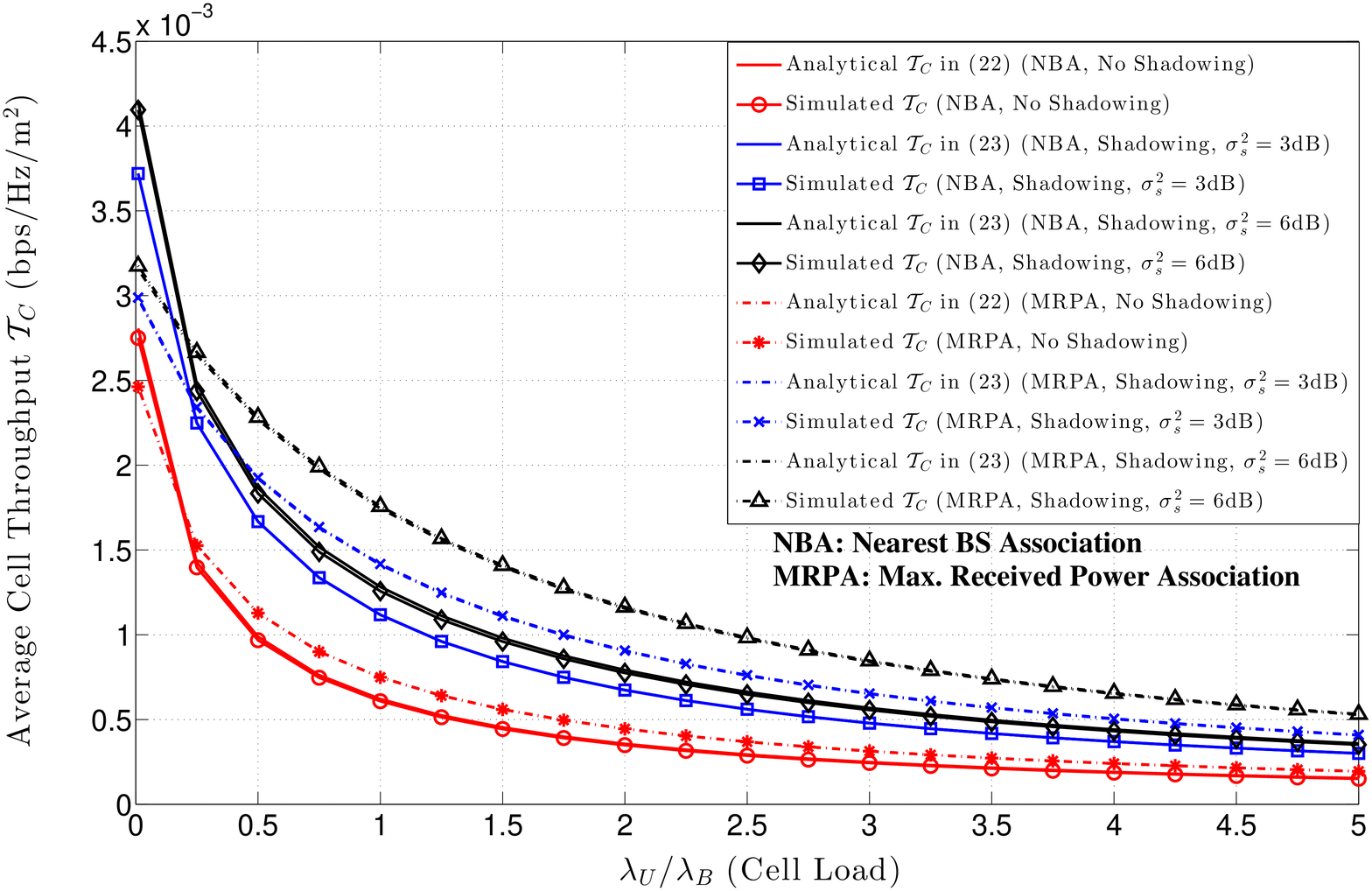}
\caption{Simulation results of average cell throughput for nearest BS association (NBA) and maximum received power association (MRPA) under Rayleigh fading and log-normal shadowing. The simulation parameters are $\alpha=3.76$, $\lambda_{\subU}=370$ users/km$^2$, $\mu_s=0$ dB and $n=6$ for eq. \eqref{Eqn:AvgCellThroughput}.}
\label{Fig:AvgCellThroughput}
\end{figure}

A numerical illustration of $\mathcal{T}_{\subC}$ for nearest BS association and maximum received power association is shown in Fig. \ref{Fig:AvgCellThroughput}. As can be seen, the analytical results in \eqref{Eqn:AvgCellThroughputNoShadow} with $n=6$ perfectly coincide with  their corresponding simulation results in different cell association schemes and shadowing conditions. All simulation curves show that $\mathcal{T}_{\subC}$ is a monotonic decreasing and convex function of the cell load for a given $\lambda_{\subU}$ and eventually it will approaches to a constant regarding that $\lambda_{\subU}$\footnote{Note that $\mathcal{T}_{\subC}$ will increase along the cell load and finally converge to a constant if the BS intensity is fixed. This is the desired situation we wish to have since $\mathcal{T}_{\subC}$ and cell load both increase at the same time.}. Most importantly, they indicate that maximum received power association evidently outperforms nearest BS association in terms of the average cell throughput since it has a larger $\zeta$, as we have already pointed out in above.  

\subsection{Average User Throughput}
Although the average cell throughput found in Proposition \ref{Prop:AvgCellThroughput} presents the per-cell spectrum efficiency of the small cell network with cell voidness, it fails to delineate how much of the downlink throughput resource in a cell can be fairly allocated to users either in the frequency or in the time domain. To characterize the average throughput fairly shared by users,  we define the following average user throughput
\begin{equation}\label{Eqn:DefnAvgUserThrou}
\mathcal{T}_{\subU} \defn \sup_{B^*_0\in\Phi_{\subB}}\left\{ \mathbb{E}\left[\log_2\left(1+\frac{H_0\|B^*_0\|^{-\alpha}}{I_0}\right)\right]\bigg/\mathbb{E}\left[\Phi_{\subU}(\mathcal{C}_0^*)\right]\right\},
\end{equation}
where $\Phi_{\subU}(\mathcal{C}_0^*)$ is the number of serviced users in a non-void cell. Note that the distribution of $\Phi_{\subU}(\mathcal{C}_0^*)$ is not equal to that of $\Phi_{\subU}(\mathcal{A})$ for any bounded Borel set $\mathcal{A}\subset\mathbb{R}^2$ with $\nu(\mathcal{A})=\nu(\mathcal{C}_0^*)$ due to the cell voidness issue. The following proposition gives the explicit result of $\mathcal{T}_{\subU}$.
\begin{proposition}
If the shadowing effect is considered in channels, the average user throughput in \eqref{Eqn:DefnAvgUserThrou} can be explicitly found as
\begin{align}\label{Eqn:AvgUserThrouShadow}
\mathcal{T}_{\subU}=\sum_{i=1}^{n} \int_{0}^{\infty}\frac{\frac{\lambda_{\subB}(1-p_{\emptyset})^2}{\lambda_{\subU}(\ln 2)}\omega_i\dif s}{[s+e^{-(\sqrt{2}\sigma_s x_i+\mu_s)}][1+(1-p_{\emptyset})\ell(s,\zeta)/\zeta]}. 
\end{align}
If the shadowing effect is not considered,  the average user throughput is
\begin{align}\label{Eqn:AvgUserThrouNoShadow}
\hspace{-0.15in}\mathcal{T}_{\subU}  =\frac{\lambda_{\subB}(1-p_{\emptyset})^2}{\lambda_{\subU}(\ln 2)}\int_{0}^{\infty} \frac{\dif s}{(s+1)[1+(1-p_{\emptyset})\ell(s,\zeta)/\zeta]}, 
\end{align}
where $\omega_i$, $x_i$, and $\ell(s,\zeta)$ are the same as given in Proposition \ref{Prop:AvgCellThroughput}.
\end{proposition}

\begin{IEEEproof}
According to the pdf of the area of associated BS $\mathcal{C}^*_0$ in \eqref{Eqn:PdfAssCell}, the  pmf pf $\Phi_{\subU}(\mathcal{C}^*_0)$ is 
\begin{align*}
&\mathbb{P}\left[\Phi_{\subU}(\mathcal{C}^*_0)=n\right]=\mathbb{E}\left[\frac{(\lambda_{\subU}\nu(\mathcal{C}^*_0))^n}{n!}e^{-\lambda_{\subU}\nu(\mathcal{C}^*_0)}\bigg| n\geq 1\right]\\
&=\frac{\lambda^n_{\subU}(\rho\lambda_{\subB}(1-p_{\emptyset}))^{\rho}}{n!(1-p_{\emptyset})\Gamma(\rho)} \int_{0}^{\infty} x^{n-1+\rho} e^{-(\lambda_{\subU}+\rho(1-p_{\emptyset})\lambda_{\subB})x}\dif x
\end{align*} 
\begin{align*}
&=\frac{p_{\emptyset}}{(1-p_{\emptyset})}\frac{\Gamma(n+\rho)}{\Gamma(\rho)n!}\left(\frac{\lambda_{\subU}/\lambda_{\subB}(1-p_{\emptyset})}{\lambda_{\subU}/\lambda_{\subB}(1-p_{\emptyset})+\rho}\right)^n,\quad n\geq 1.
\end{align*}
where $p_{\emptyset}$ is given in \eqref{Eqn:VoidProbGCA}. As a result, the mean number of users in cell $\mathcal{C}^*_0$ is
\begin{align*}
\mathbb{E}\left[\Phi_{\subU}(\mathcal{C}^*_0)\right]&=\frac{p_{\emptyset}}{(1-p_{\emptyset})\Gamma(\rho)}\times\\ &\hspace{0.2in}\sum_{n=1}^{\infty}\frac{\Gamma(n+\rho)}{(n-1)!}\left(\frac{\lambda_{\subU}/\lambda_{\subB}(1-p_{\emptyset})}{\lambda_{\subU}/\lambda_{\subB}(1-p_{\emptyset})+\rho}\right)^n\\
&=\frac{\lambda_{\subU}}{\lambda_{\subB}(1-p_{\emptyset})^2}.
\end{align*}
The average user throughput can be calculated as follows
\begin{align}
\mathcal{T}_{\subU}=\frac{\lambda_{\subB}(1-p_{\emptyset})^2\mathcal{T}_{\subC}}{\lambda_{\subU}(\mathbb{E}[1/\nu(\mathcal{C}^*_0)])}= \frac{(\rho-1)(1-p_{\emptyset})\mathcal{T}_{\subC}}{\rho\lambda_{\subU}} \label{Eqn:ProofAvgUserThro}.
\end{align}
Substituting \eqref{Eqn:AvgCellThroughput} into \eqref{Eqn:ProofAvgUserThro} results in \eqref{Eqn:AvgUserThrouShadow} and \eqref{Eqn:AvgUserThrouNoShadow}. 
\end{IEEEproof}

The average user throughput in \eqref{Eqn:AvgUserThrouShadow} and \eqref{Eqn:AvgUserThrouNoShadow} is  a linear scaling of the average cell throughput normalized by the user intensity such that it is virtually the function of cell load $\lambda_{\subU}/\lambda_{\subB}$. For the asymptotic behaviors of $\mathcal{T}_{\subU}$, we have $\mathcal{T}_{\subU}\in\Theta\left(\frac{\lambda_{\subB}}{\lambda_{\subU}(1+\ell(s,\zeta)/\zeta)}\right)$ and $\mathcal{T}_{\subU}\rightarrow 0$ in  the regime of high $\lambda_{\subU}/\lambda_{\subB}$ (i.e., the regime of heavy cell load) since there are too many users sharing the throughput resource in a cell.  On the other hand, in the regime of low $\lambda_{\subU}/\lambda_{\subB}$ (i.e., the regime of light cell load),  $\mathcal{T}_{\subU}$ scales like $\Theta(\lambda_{\subU}/\lambda_{\subB})$ and $\mathcal{T}_{\subU}\rightarrow 0$ since  the average area of non-void cells is so large that the average throughput shared by users is fairly small.  Specifically, in moderate cell load  there exists a sole optimal cell load that maximizes the average user throughput. Knowing this optimal cell load is important since it indicates when cell load and average user throughput increase at the same time and  in this case more user throughput can be exploited by deploying \textit{fewer} BSs for a given user intensity. The following proposition gives the accurate solution of the optimal cell load for $\mathcal{T}_{\subU}$. 
\begin{proposition}\label{Prop:OptCellLoadUserThr}
The optimal cell load that maximizes the average user throughputs in \eqref{Eqn:AvgUserThrouShadow} and \eqref{Eqn:AvgUserThrouNoShadow} is the fixed point of the following real-valued function parameterized by $\beta>1$
\begin{equation}\label{Eqn:OptCellLoadCond}
\mathsf{L}(v) =\left[1-\left(1+\frac{v}{\rho}\right)^{-\rho}\right]^{\frac{1}{\beta}},\quad v\geq 0.
\end{equation} 
Namely, if the optimal cell load is $v^*$, we have $\mathsf{L}(v^*)=v^*$.
\end{proposition}
\begin{IEEEproof}
Since the cell load is only contained in the term $(1-p_{\emptyset})^2\lambda_{\subB}/\lambda_{\subU}[1+(1-p_{\emptyset})\ell(s,\zeta)/\zeta]$, we find its optimal value by merely calculating the derivatives of this term with respective to the cell load. Let $v$ denote the cell load,  $\mathcal{V}(v)$ represent this term parameterized by $v$and $\tilde{p}_{\emptyset}=1-p_{\emptyset}$. Then we have
$$\mathcal{V}(v) = \frac{\tilde{p}^2_{\emptyset}}{v[1+\tilde{p}_{\emptyset}\ell(s,\zeta)/\zeta]}.$$
Note that $\ell(s,\zeta)/\zeta$ is not function of $v$. Hence, the first-order derivative of $\mathcal{V}(v)$ w.r.t. $v$ is
$$\frac{\partial\mathcal{V}}{\partial v}=\frac{\tilde{p}_{\emptyset}\tilde{p}'_{\emptyset}v[2+\tilde{p}_{\emptyset}\ell(s,\zeta)/\zeta]-\tilde{p}^2_{\emptyset}[1+\tilde{p}_{\emptyset}\ell(s,\zeta)/\zeta]}{v^2[1+\tilde{p}_{\emptyset}\ell(s,\zeta)/\zeta]^2},$$
where
$$\tilde{p}'_{\emptyset}=-\frac{\dif p_{\emptyset}}{\dif v}=\left(1+\frac{v}{\rho }\right)^{-\rho-1}=p^{1+\frac{1}{\rho}}_{\emptyset}\geq 0.$$

Letting $\frac{\partial\mathcal{V}}{\partial v}=0$ yields the following differential equation of $\tilde{p}_{\emptyset}$ and its solution
$$\frac{\dif \tilde{p}_{\emptyset}}{\dif v}=\frac{\tilde{p}_{\emptyset}[1+\tilde{p}_{\emptyset}\ell(s,\zeta)/\zeta]}{v[2+\tilde{p}_{\emptyset}\ell(s,\zeta)/\zeta]} \Rightarrow \frac{\tilde{p}^2_{\emptyset}}{1+\tilde{p}_{\emptyset}\ell(s,\zeta)/\zeta}=c(\tilde{p}_{\emptyset})v,$$
where $c$ is an unknown function of $\tilde{p}_{\emptyset}$ that can be determined as follows. First, we can assign $c(v,\tilde{p}_{\emptyset})=\tilde{p}_{\emptyset}^{2-1/\beta}/(1+\tilde{p}_{\emptyset}\ell(s,\zeta)/\zeta)$ for all $\beta\in\mathbb{R}_{++}$ since $\tilde{p}_{\emptyset}$ should only depend on $v$, not on $\ell(s,\zeta)/\zeta$. Thereby, we acquire $\tilde{p}^{1/\beta}_{\emptyset}=v$. Since  $\tilde{p}'_{\emptyset}=-p'_{\emptyset}=p^{1+\frac{1}{\rho}}_{\emptyset}$, we know $\tilde{p}'_{\emptyset}\in[0,1]$ and $p'_{\emptyset}<0$. To make $\tilde{p}^{1/\beta}_{\emptyset}(v)$ have a fixed point at $v^*$, the first-order and second-order derivative constraints on $\tilde{p}^{1/\beta}_{\emptyset}(v)$,  $\frac{1}{\beta}\tilde{p}^{1/\beta-1}_{\emptyset}(0)\tilde{p}'_{\emptyset}(0)> 0$ and $\frac{1}{\beta}(\frac{1}{\beta}-1)\tilde{p}^{1/\beta-2}_{\emptyset}(\tilde{p}'_{\emptyset})^2+\frac{1}{\beta}\tilde{p}^{1/\beta-1}_{\emptyset}\tilde{p}''_{\emptyset}< 0$ for all $v>0$, must hold. Since they hold only when $\beta >1$, $\mathsf{L}(v)$ in \eqref{Eqn:OptCellLoadCond} is obtained. Finally, we can show
$$\lim_{v\rightarrow 0}\frac{\partial\mathcal{V}}{\partial v}>0\quad\text{and}\quad \lim_{v\rightarrow\infty}\frac{\partial\mathcal{V}}{\partial v}<0,$$
and the solution of $\tilde{p}^{1/\beta}_{\emptyset}(v^*)-v^*=0$ for $v^*\geq 0$ is unique since $v^*\geq 0$ and $\tilde{p}_{\emptyset}$ is a positive and one-to-one mapping function of $v$. This follows that $\mathsf{L}(v^*)=(1-p_{\emptyset})^{1/\beta}=\tilde{p}^{1/\beta}_{\emptyset}=v^*$ and the fixed point of $\mathsf{L}(v^*)$ is the  optimal cell load that maximizes $\mathcal{T}_{\subU}$.
\end{IEEEproof}
\begin{remark}
For any GCA schemes with $\rho\gg 1$, \eqref{Eqn:OptCellLoadCond} reduces to $\mathsf{L}(v) \approx (1-\exp(-v))^{1/\beta}$ whose fixed point is $v=0$. Hence, the optimal cell load asymptotically reduces to zero as $\rho$ goes to infinity.   
\end{remark}

\begin{figure}[t!]
\centering
\includegraphics[width=3.65in,height=2.6in]{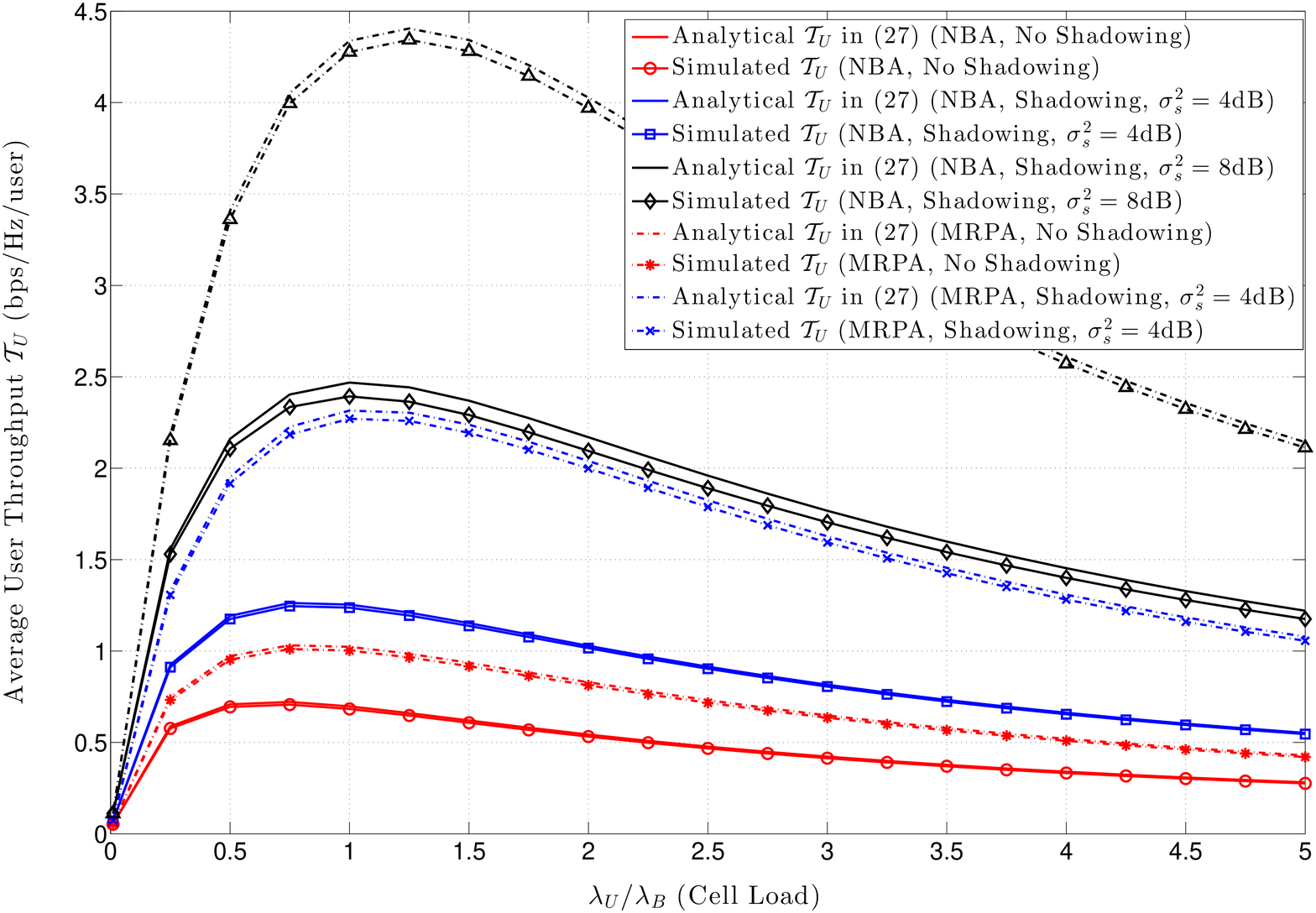}
\caption{Simulation results of average user throughput for nearest BS association (NBA) and maximum received power association (MRPA) under Rayleigh fading and log-normal shadowing. The simulation parameters are $\alpha=3.76$, $\mu_s=0$ dB, $\lambda_{\subU}=370$ users/km$^2$ and $n=6$ for eq. (27).}
\label{Fig:AvgUserThroughput}
\end{figure}

Cell association also has a pivotal influence on the average user throughput same as the case in the average cell throughput, and this situation can be visually perceived from the numerical results of nearest BS and maximum received power associations in Fig. \ref{Fig:AvgUserThroughput}. Indeed, large shadowing power apparently increases $\mathcal{T}_{\subU}$ as well. There exists only one optimal cell load that maximizes $\mathcal{T}_{\subU}$ as shown in Proposition \ref{Prop:OptCellLoadUserThr}. For example, in the case of maximum received power association with $\alpha=3.76$ and shadowing variance $\sigma^2_s=4$ dB, Proposition \ref{Prop:OptCellLoadUserThr} delivers an optimal cell load $\approx 1$ for $\rho=10.3$ and $\beta=2$, which accurately coincides with the optimal cell load observed in the corresponding curve case of Fig. \ref{Fig:AvgUserThroughput}. In addition, we also observe that the optimal cell load happens later as shadowing power gets larger, which further implies that shadowing is instructive to the efficiency of deploying cells since higher average user throughput is attained by less deployment cost.  

\section{Optimal Green Cell Load and Throughput}\label{Sec:OptGreenIntesityThrou}
The average cell and user throughputs found in the previous section have a common drawback, that is, the energy cost is not integrated into them while considering to achieve higher throughput by deploying more cells. Deploying more cells may not necessarily leads to more total power consumption since the transmit power in \eqref{Eqn:ReqTxPower} (that indicates $P_t\propto \lambda^{-\frac{\alpha}{2}}_{\subB}$) decreases due to less path loss even though other hardware power consumptions increase in this case. In order to delve the fundamental relationship between throughput and power consumption, we define the following green cell throughput $\mathcal{G}_{\subC}$ (bits/Hz/joule) and green user throughput $\mathcal{G}_{\subU}$ (bits/Hz/joule/user):
\begin{equation}\label{Eqn:DefnGreenThroughput}
\mathcal{G}_{\subC} =\frac{\mathcal{T}_{\subC}}{\lambda_{\subB}\mathbb{E}[\Psi]} \,\, \text{ and }\,\, \mathcal{G}_{\subU} =\frac{\mathcal{T}_{\subU}}{\mathbb{E}[\Psi]} 
\end{equation}
in which $\mathbb{E}[\Psi]$ is the average power consumption of a BS and it is already given in \eqref{Eqn:AvgComPower}. The physical meanings of green cell and user throughputs are how much throughput can be transported by costing unit power consumption. Note that the definitions of $\mathcal{G}_{\subC}$ and $\mathcal{G}_{\subU}$ are slightly different since $\mathcal{G}_{\subC}$ aims to quantify an energy efficiency on the throughput of a BS whereas $\mathcal{G}_{\subU}$ desires to characterize an energy efficiency on the throughput of a user. In this section, we would like to investigate the fundamental relationship between cell load and these two green throughputs in order to gain some insights on how to deploy BSs with an appropriate intensity for achieving high throughput efficiency. The green cell and user throughputs are studied in Sections \ref{SubSec:OptGreenCellLoad} and \ref{SubSec:OptGreenUserLoad}, respectively.

\subsection{Green Cell Throughput and Its Optimality}\label{SubSec:OptGreenCellLoad}
According to \eqref{Eqn:DefnGreenThroughput} and \eqref{Eqn:AvgCellThroughput}, the green cell throughput can be  specifically expressed as
\begin{equation}\label{Eqn:GreenCellEngThrou}
\mathcal{G}_{\subC}=\dfrac{\sum_{i=1}^{n}\int_{0}^{\infty}\frac{\dif s}{[s+e^{-(\sqrt{2}\sigma_sx_i+\mu_s)}][1+(1-p_{\emptyset})\ell(s,\zeta)/\zeta]}}{(\ln 2)(1-1/\rho)[(P_{\subON}+\delta P_t)+p_{\emptyset}P_{\subOFF}/(1-p_{\emptyset})]},
\end{equation}
where $P_t=\frac{P_{\min}\Gamma(1+\frac{\alpha}{2})}{(\pi\lambda_{\subB}\zeta)^{\frac{\alpha}{2}}}$ as already given in \eqref{Eqn:ReqTxPower}. In heavy cell load, $p_{\emptyset}\approx 0$ and $\mathcal{G}_{\subC}$ approaches to a constant approximated by
$$\mathcal{G}_{\subC}\approx \frac{\sum_{i=1}^{n}\int_{0}^{\infty}\frac{\dif s}{(s+e^{-(\sqrt{2}\sigma_sx_i+\mu_s)})(1+\ell(s,\zeta)/\zeta)}}{(\ln 2)(1-1/\rho)(P_{\subON}+\delta P_t)},$$
which can be viewed as the ``least'' green cell throughput merely affected by GCA and BS intensity since the increase in the power consumption of the active mode has a direct impact on $\mathcal{G}_{\subC}$. In this case, deploying more BSs and using appropriate cell association schemes with a larger $\zeta$ can make cell throughput greener.  In light cell load, $p_{\emptyset}\approx 1$ and $\mathcal{G}_{\subC}$ has the following approximation
$$ \mathcal{G}_{\subC}\approx \left(\frac{\lambda_{\subU}}{\lambda_{\subB}}\right) \frac{\sum_{i=1}^{n}\int_{0}^{\infty}\frac{\dif s}{[s+e^{-(\sqrt{2}\sigma_sx_i+\mu_s)}]}}{(\ln 2)(1-1/\rho)P_{\subOFF}},$$
i.e., $\mathcal{G}_{\subC}=\Theta(\lambda_{\subU}/\lambda_{\subB}P_{\subOFF})$ as $\lambda_{\subU}/\lambda_{\subB}\rightarrow 0$, and it is the greenest cell throughput since any additional power consumption in the active mode hardly reduces the cell throughput. However, $\mathcal{G}_{\subC}$ in this case would be very low since it is dominated by the light cell load. Lowering the power consumption of the dormant mode and increasing cell load can linearly augment the green cell throughput. 

The two aforementioned extreme cases indicate that the cell load should not be too heavy or too light in order to maintain a certain level of cell throughput with satisfactory greenness. The exists a unique optimal cell load that maximizes the green cell throughput for a given user intensity as shown in the following proposition.
\begin{proposition}\label{Prop:OptGreenCellLoad}
For a given user intensity $\lambda_{\subU}$, there exists a $\beta>1$ such that the unique optimal cell load that maximizes $\mathcal{G}_{\subC}$ in \eqref{Eqn:GreenCellEngThrou} can be accurately found by the fixed point of the following real-valued function
\begin{align}\label{Eqn:OptGreenCellLoad}
\mathds{L}_{\subC}(v)=&\pi\zeta\lambda_{\subU}\Bigg(\frac{(P_{\subON}-P_{\subOFF})}{\delta P_{\min}\Gamma(1+\frac{2}{\alpha})}\nonumber\\
&\left\{\left[1-\left(1+\frac{v}{\rho}\right)^{-\rho}\right]^{-\frac{1}{\beta}}-1\right\}\Bigg)^{\frac{2}{\alpha}}.
\end{align}
Namely, the optimal cell load is the unique solution of $\mathds{L}_{\subC}(v^*)-v^*=0$ such that the corresponding optimal BS intensity is $\lambda^*_{\subB}=\lambda_{\subU}/v^*$.
\end{proposition}
\begin{IEEEproof}
See Appendix \ref{App:OptGreenCellLoad}.
\end{IEEEproof}

There are some interesting implications that can be perceived from the fixed point result in \eqref{Eqn:OptGreenCellLoad}. First of all, a nonzero optimal cell load does not exist if BSs do not perform green power control, i.e., $P_{\subON}=P_{\subOFF}$, whereas the more power is saved (i.e., $P_{\subON}-P_{\subOFF}$ is larger), the larger optimal cell load and green cell throughput are achieved. Cell association schemes with large $\zeta$ increase the optimal cell load as well as the optimal green throughput at the same time. This is because large $\zeta$ reduces the void probability, yet increases the coverage probability. Furthermore, cell association has a more direct impact on the optimal cell load than green power control since the effect of green power control is weakened by the power of $\frac{2}{\alpha}$. Although green power control and cell association both benefit the optimal cell load and green cell  throughput, green power seems more favorable in practice since it is easily implemented and not dependable on the communication environments.

\subsection{Green User Throughput and Its Optimality}\label{SubSec:OptGreenUserLoad}
 Substituting \eqref{Eqn:AvgUserThrouShadow} into \eqref{Eqn:DefnGreenThroughput} results in the explicit expression of the green user throughput  given by
\begin{equation}\label{Eqn:GreenUserEngThrou}
\mathcal{G}_{\subU}=\dfrac{\lambda_{\subB}(1-p_{\emptyset})\sum_{i=1}^{n}\int_{0}^{\infty}\frac{\dif s}{[s+e^{-(\sqrt{2}\mu_i+\sigma_i)}][1+(1-p_{\emptyset})\ell(s,\zeta)/\zeta]}}{(\ln 2)\lambda_{\subU}[(P_{\subON}+\delta P_t)+p_{\emptyset}P_{\subOFF}/(1-p_{\emptyset})]}.
\end{equation}
The asymptotic characteristics of $\mathcal{G}_{\subU}$ are different from those of $\mathcal{G}_{\subC}$. In heavy cell load, we have the scaling law of the ``least'' green user throughput, i.e.,   $\Theta(\lambda_{\subB}/\lambda_{\subU}(P_{\subON}+\delta P_t))$ as $p_{\emptyset}\rightarrow 0$, which is dominated by the power consumption of the active mode and cell load.  From the green user throughput point of view, heavy cell load should be avoided since any additional power consumption in the active mode completely gives rise to user throughput reduction. In light cell load, we have  $\mathcal{G}_{\subU}=\Theta(\lambda_{\subU}/\lambda_{\subB}P_{\subOFF})$ as $p_{\emptyset}\rightarrow 1$, which is the same as the scaling law of the greenest cell throughput and also suffers the low throughput problem while achieving high greenness. Light cell load should also be dodged from the perspective of user's quality of service. 

Since too heavy and too light cell loads do not benefit the green user throughput, we resort to finding the optimal cell load for $\mathcal{G}_{\subU}$ if possible.
Indeed, the following proposition shows that $\mathcal{G}_{\subU}$ also has a sole maximizer of cell load and how to find it. 
\begin{proposition}\label{Prop:OptGreenUserLoad}
For a given user intensity $\lambda_{\subU}$, there exists a $\beta>1$ such that the optimal cell load that maximizes $\mathcal{G}_{\subU}$ in \eqref{Eqn:GreenCellEngThrou} can be accurately found by the sole fixed point of the following real-valued function
\begin{align}\label{Eqn:OptGreenUserLoad}
\mathds{L}_{\subU}(v)=&\pi\zeta\lambda_{\subU}\Bigg(\frac{(P_{\subON}-P_{\subOFF})}{\delta P_{\min}\Gamma(1+\frac{2}{\alpha})}\nonumber\\
&\left\{\left[1-\left(1+\frac{v}{\rho}\right)^{-\rho}\right]^{-\frac{2}{\beta}}-1\right\}\Bigg)^{\frac{2}{\alpha+\beta}},
\end{align}
which means $v^*$ is the unique maximizer of cell load for $\mathcal{G}_{\subU}$ if and only if $\mathds{L}_{\subU}(v^*)-v^*=0$ and the optimal BS intensity is $\lambda^*_{\subB}=\lambda_{\subU}/v^*$.
\end{proposition}
\begin{IEEEproof}
The proof is omitted here since it is similar to the proof of Proposition \ref{Prop:OptGreenCellLoad}.
\end{IEEEproof}

Obviously, the function in \eqref{Eqn:OptGreenUserLoad} for the optimal cell load of the green user throughput is similar to the function in \eqref{Eqn:OptGreenCellLoad} for the optimal cell load of the green cell throughput. Its fixed point also becomes larger if cell association schemes have a larger $\zeta$. Under the same conditions, since green cell throughput is essentially smaller than green cell throughput and the increasing rates of $\mathcal{G}_{\subC}$ and $\mathcal{G}_{\subU}$ w.r.t. the cell load are pretty much the same in the light cell load, the optimal cell load for the green user throughput tends to appear earlier than that for the optimal cell throughput. This phenomenon can be observed from Proposition \ref{Prop:OptGreenCellLoad} as well as the simulation results in the following subsection.

\subsection{Simulation Results}
This subsection presents some simulation results to verify whether or not the previous results and discussions regarding the green cell and user throughputs are correct. We adopt the urban micro path loss model for 2GHz carrier frequency ($140.7 + 37.6\log_{10}d,\, d$ in km) given in \cite{3GPP36.814.10} incorporated with average penetration loss (20 dB for 80\% users that are indoor and 0 dB for the rest that are outdoor), antenna gain 5 dBi\cite{3GPP36.872.13}. The power consumption model of the small cell BS is based on the model of picocells  in \cite{GAVGCD11}. We summarize the common network parameters for all simulation figures as follows: $P_{\subON} = 6.8$W, $P_{\subOFF}=4.3$W, $P_{\min}=-106$ dBm, $\delta = 4.0$, $\mu_s=0$ dB and $\alpha=3.76$. For simplicity, only two GCA schemes, i.e., nearest BS association and maximum received power association, are used to do simulation and compare the simulation results.  

\begin{figure}[t!]
\centering
\includegraphics[width=3.65in,height=2.6in]{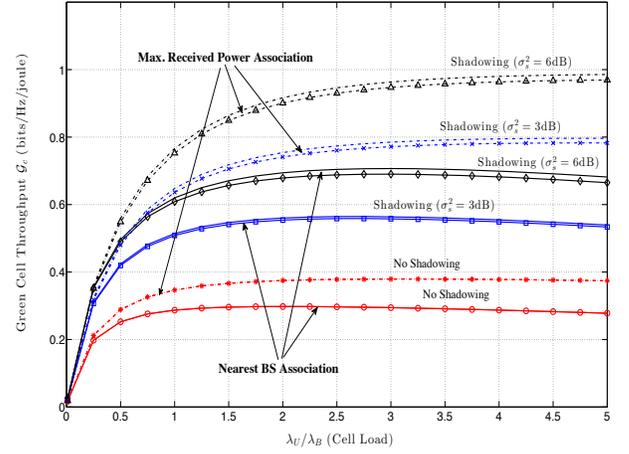}
\caption{Simulation results of green cell throughput for nearest BS association and maximum received power association under Rayleigh fading and log-normal shadowing. The user intensity is $\lambda_{\subU}=370$ users/km$^2$ and $n=6$ is used for eq. (31).}
\label{Fig:GreenCellThroughput}
\end{figure}

\begin{figure}[t!]
\centering
\includegraphics[width=3.65in,height=2.6in]{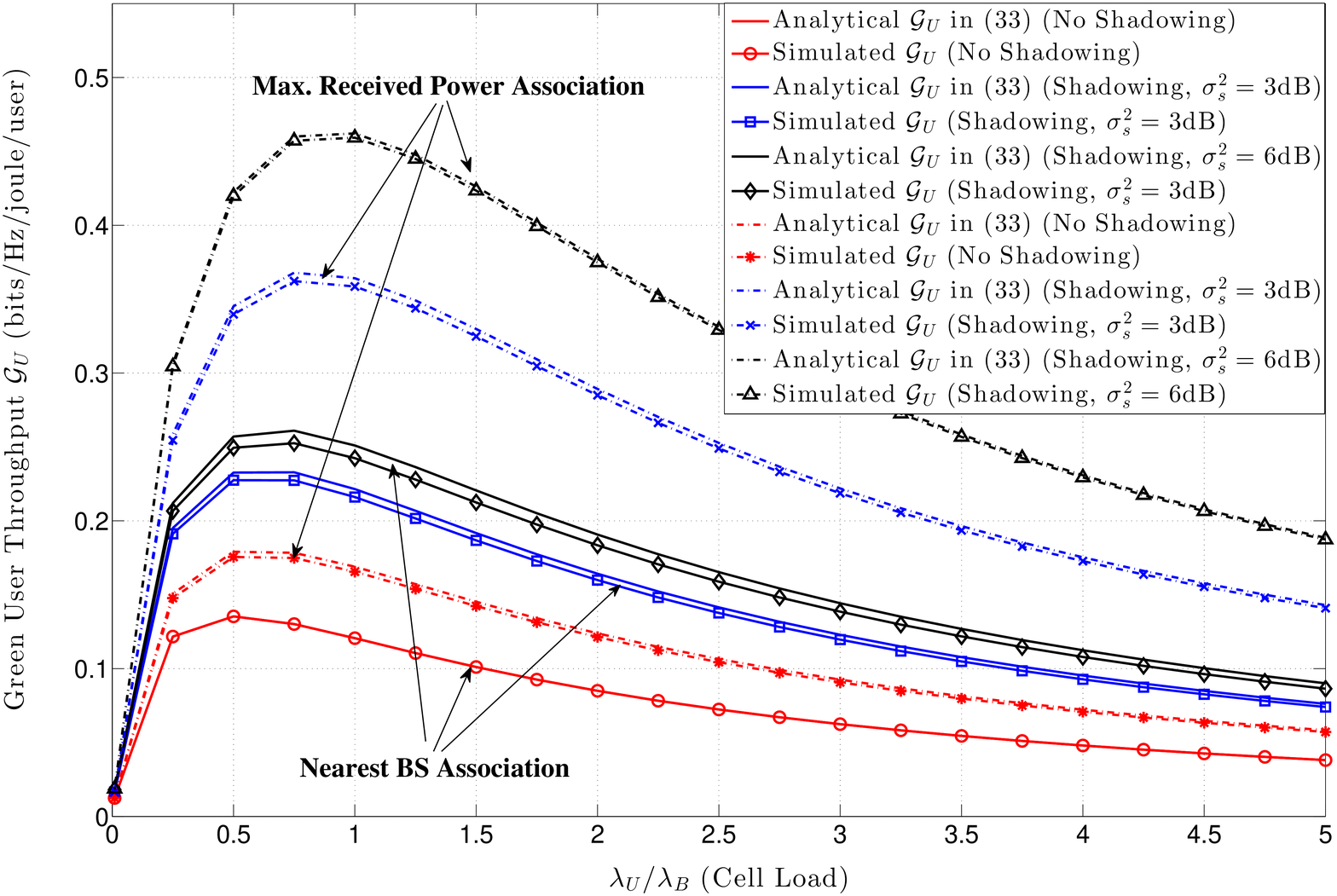}
\caption{Simulation results of green user throughput for nearest BS association and maximum received power association under Rayleigh fading and log-normal shadowing. The user intensity is $\lambda_{\subU}=370$ users/km$^2$ and $n=6$ is used for eq. (33).}
\label{Fig:GreenUserThroughput}
\end{figure}

\begin{figure}[t!]
	\centering
	\includegraphics[width=3.65in,height=2.6in]{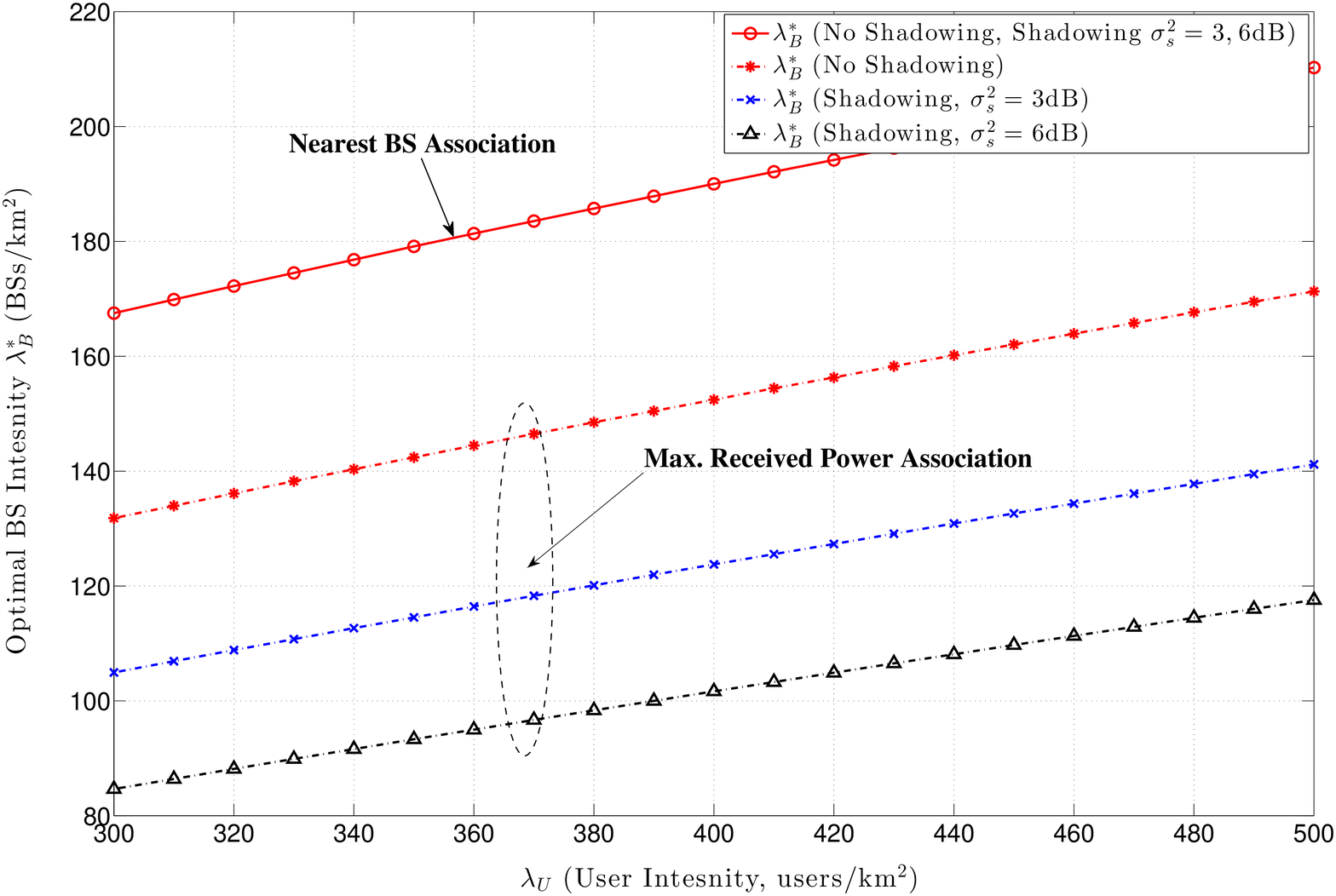}
	\caption{Simulation results of the optimal BS intensity for green cell throughput under different user intensities.}
	\label{Fig:OptBSIntenGreenCellThrou}
\end{figure}

\begin{figure}[t!]
	\centering
	\includegraphics[width=3.65in,height=2.6in]{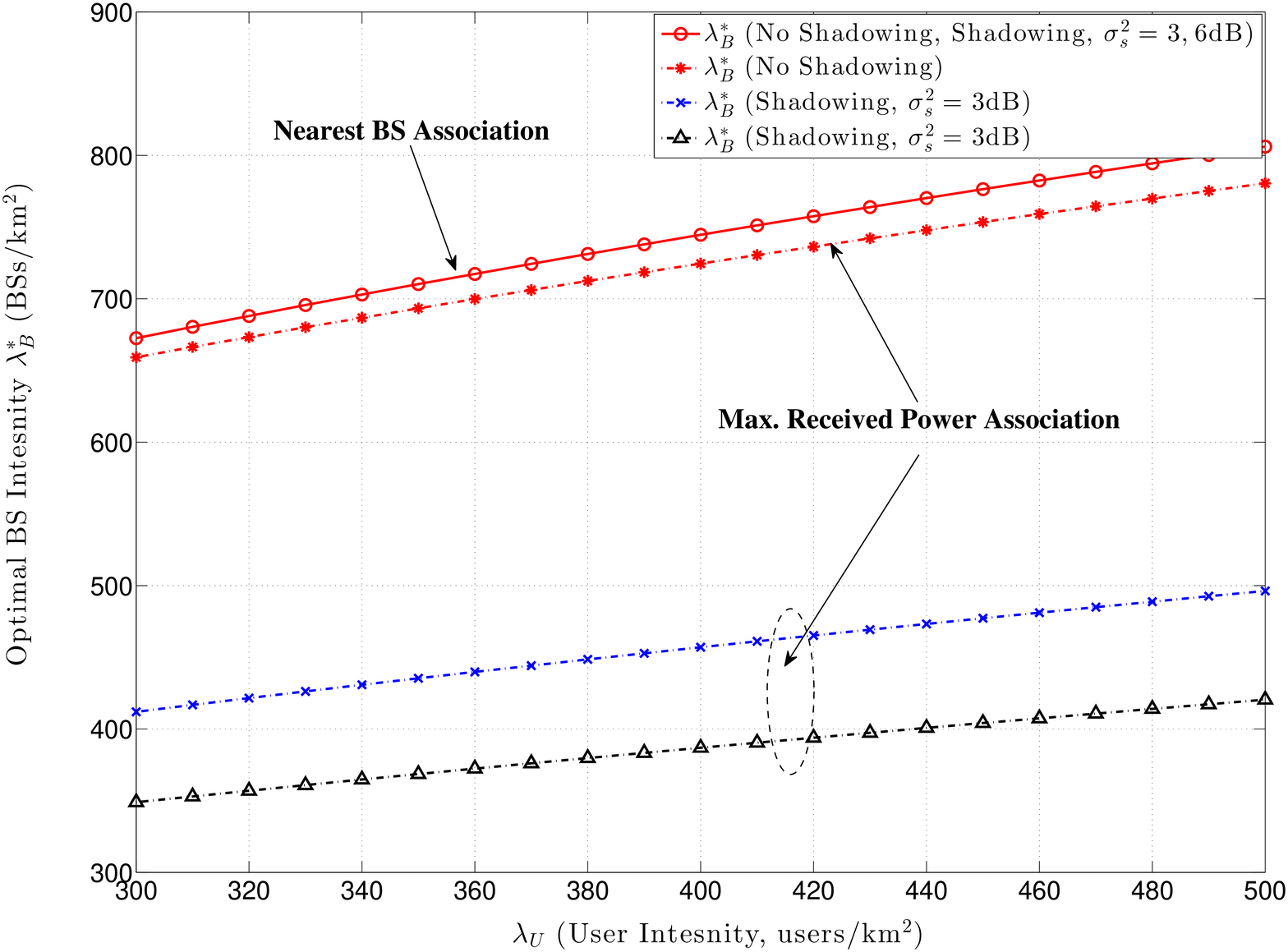}
	\caption{Simulation results of the optimal BS intensity for green user throughput under different user intensities.}
	\label{Fig:OptBSIntenGreenUserThrou}
\end{figure}

Figs. \ref{Fig:GreenCellThroughput} and \ref{Fig:GreenUserThroughput} show the simulation results of the green cell and user throughputs, respectively. Both figures reveal that maximum received power association outperforms nearest BS association, shadowing power helps improve the green cell and user throughputs, and only one optimal cell load exists for every different simulation cases, as expected. Also, stronger shadowing power lets the optimal cell load happen later. To further explain this point, the simulation results of the optimal BS intensities for green cell and user throughputs versus different user intensities are
provided in Figs. \ref{Fig:OptBSIntenGreenCellThrou} and \ref{Fig:OptBSIntenGreenUserThrou}, respectively. From these two figures, we see that the slopes of the green cell and user throughputs are initially both positive and reduce largely as the shadowing power increases (see the previous discussion for $\mathcal{G}_{\subC}$ in light cell load), which is because large shadowing power still has a strong influence on the coverage probability in \eqref{Eqn:LapInter} but has very less impact on the void cell probability. Propositions \ref{Prop:OptGreenCellLoad} and \ref{Prop:OptGreenUserLoad} also indicate this phenomenon and thus stronger shadowing power indeed leads to a smaller optimal BS intensity and larger green cell and user throughputs for the same user intensity, which is desirable. Finally, we observe that optimal BS intensities found by \eqref{Eqn:OptGreenCellLoad} with $\beta\in(6,8)$ and \eqref{Eqn:OptGreenUserLoad}  with $\beta\in(1,4)$ perfectly coincide with the numerical results in Figs. \ref{Fig:OptBSIntenGreenCellThrou} and \ref{Fig:OptBSIntenGreenUserThrou}. The correctness of Propositions \ref{Prop:OptGreenCellLoad} and \ref{Prop:OptGreenUserLoad} is validated.  

\section{Conclusion}\label{Sec:Conclusion}
 In this paper, we investigate the optimal cell load, average and green throughputs from the cell and user perspectives in a small cell network with cell voidness.  The impacts from the void cell phenomenon are two-fold. Firstly, it indicates that the previous works on throughput or rate are no longer accurate especially in a small cell network with dense BS deployment since no interferences are generated by the void cells. Secondly, it inspires the idea of green power control that makes the void BSs switch to the dormant mode for saving power.  An accurate expression of the void cell probability is successfully derived for GCA. The average cell and user throughputs that reflect the impact of void cells are theoretically characterized and they can be effectively boosted if appropriate cell load and GCA schemes with strong channel power (such as fading and shadowing) are adopted. The proposed green power control scheme not only exploits power saving on the void BSs but also gives birth to the idea of defining the green cell and user throughputs on the basis of the average cell and user throughputs. The green cell and user throughputs can be maximized by their unique optimal cell load that can be significantly improved by green power control and GCA.

\appendix
\subsection{Proof of Theorem \ref{Thm:RanTransProp}}\label{App:ProofRanTransProp}
The following proof essentially generalizes the proof of Lemma 1 in \cite{CHLBRSC15}. First, we define the void probability of $\Phi'\cap\mathcal{A}'$ for any bounded Borel set $\mathcal{A}'\subset\mathbb{R}^{\mathrm{d}'}$ as the probability that $\Phi'(\mathcal{A}')$ is equal to zero where $\Phi'(\mathcal{A}')$ denotes the number of nodes in $\Phi'$ enclosed in $\mathcal{A}'$. That is,
$$\mathbb{P}\left[\Phi'(\mathcal{A}')=0\right] =\exp\left(-\int_{\mathcal{A}'} \Lambda'(\dif X')\right)$$
because $\Phi'$ is a PPP. Let $\mathcal{A}_i=\mathbf{T}_i(\mathcal{A}')$ where $\mathcal{A}_i\subset\mathbb{R}^{\mathrm{d}}$ is a bounded Borel set and the probability that node $\hat{X}_i$ in $\hat{\Phi}$ is enclosed in $\mathcal{A}_i$ is
\begin{align*}
\mathbb{P}\left[\hat{X}_i\in\mathcal{A}_i\right]&=\mathbb{P}\left[\mathbf{T}_i(X'_i)\in\mathcal{A}_i \right] = \mathbb{P}\left[\mathbf{T}^{\mathrm{T}}_i\mathbf{T}_i(X'_i)\in \mathbf{T}^{\mathrm{T}}_i(\mathcal{A}_i) \right]\\ 
&= \mathbb{P}\left[X'_i \in(\mathbf{T}^{\mathrm{T}}_i\mathbf{T}_i)^{-1}(\mathcal{A}') \right],
\end{align*}
where $(\mathbf{T}_i^{\mathrm{T}}\mathbf{T}_i)^{-1}(\mathcal{A}') \defn \int_{\mathcal{A}'}(\mathbf{T}_i^{\mathrm{T}}\mathbf{T}_i)^{-1}(\dif X')$.  By definition, $\hat{\Lambda}(\mathcal{A})$ is the average of $\hat{\Phi}(\mathcal{A})$ and it can be calculated by applying the Campbell theorem as shown in the following:
\begin{align*}
\hat{\Lambda}(\mathcal{A})&=\int_{\mathcal{A}}\hat{\Lambda}(\dif \hat{X}) =\mathbb{E}\left[\sum_{\hat{X}_i\in\hat{\Phi}}\mathds{1}\left(\hat{X}_i\in\hat{\Phi}\cap\mathcal{A}_i\right)\right]\\
&=\mathbb{E}\left[\sum_{X_i\in\Phi'}\mathds{1}\left(X'_i\in\Phi'\cap (\mathbf{T}_i^{\mathrm{T}}\mathbf{T}_i)^{-1}(\mathcal{A}')\right)\right]\\
&=\mathbb{E}\left[ \int_{ (\mathbf{T}^{\mathrm{T}}\mathbf{T})^{-1}(\mathcal{A}') }\mathbb{P}\left[X'\in\Phi' \right] \nu_{\mathrm{d}'}(\dif X')\right]
\end{align*}
\begin{align*}
&\stackrel{(\star)}{=}\mathbb{E}\left[\frac{1}{\sqrt{\det(\mathbf{T}^{\mathrm{T}}\mathbf{T})}}\right]\int_{\mathbb{R}^{\mathrm{d}'}}\mathbb{P}\left[X'\in\Phi'\cap\mathcal{A}' \right] \nu_{\mathrm{d}}(\dif X')\\
&=\mathbb{E}\left[\frac{1}{\sqrt{\det(\mathbf{T}^{\mathrm{T}}\mathbf{T})}}\right] \int_{\mathcal{A}'} \Lambda'(\dif X')
\end{align*}
where $(\star)$ follows from the Jacobian determinant of two volumes. Thus,  $\hat{\Lambda}(\mathcal{A})=\mathbb{E}\left[\frac{1}{\sqrt{\det(\mathbf{T}^{\mathrm{T}}\mathbf{T})}}\right]\Lambda'(\mathcal{A}')$ and it follows that
\begin{align*}
\mathbb{P}\left[\Phi'(\mathcal{A}')=0\right]&=\exp\left(-\mathbb{E}\left[\frac{1}{\sqrt{\det(\mathbf{T}^{\mathrm{T}}\mathbf{T})}}\right]^{-1}\int_{\mathcal{A}}\hat{\Lambda}(\dif \hat{X})\right)\\
&=\mathbb{P}\left[\hat{\Phi}(\mathcal{A})=0\right]
\end{align*}
since $\nu_{\mathrm{d}}(\mathcal{A})=\nu_{\mathrm{d}'}(\mathcal{A}')$. Since the void probability of a point process completely characterizes the statistics of the process, $\hat{\Phi}$ is also a PPP.

 If $\Phi'$ is homogeneous, then its intensity measure is $\Lambda' (\mathcal{A}')=\lambda'\nu_{\mathrm{d}'}(\mathcal{A}')$ and thus the result in above can be further simplified to
 $\hat{\Lambda}(\mathcal{A})=\lambda'\mathbb{E}\left[ \sqrt{\det((\mathbf{T}^{\mathrm{T}}\mathbf{T})^{-1})}\right]\nu_{\mathrm{d}}(\mathcal{A})$ and it does not depend on the location of any node in $\hat{\Phi}$. As a result, $\hat{\Phi}$ is a homogeneous PPP with the intensity given in \eqref{Eqn:TransHomoIntensityMeasure}.

\subsection{Proof of Theorem \ref{Thm:BoundVoidProb}}\label{App:ProofBoundVoidProb} 
 Suppose each user in cell $\mathcal{C}_l\subset\mathbb{R}^2$ of BS $B_l$ has its own cell association region which is also a bounded Borel set on $\mathbb{R}^2$. For example, the cell association region of user $U_j$ is denoted by $\mathcal{A}_j$ and all cell association regions contain  BS $B_l$. Since all users in $\mathcal{C}_l$ adopt the GCA scheme, the probability that BS $B_l$ is not associated by user $U_j\in\mathcal{C}_l$ can be expressed as
\begin{align*}
 \mathbb{P}[\mathcal{E}_{l,j}]=&\lim_{\nu(\mathcal{A}_j)\rightarrow\infty}\mathbb{P}\bigg[\sup_{B_i\in\Phi_{\subB}\cap\mathcal{A}_j\setminus B_l}\left\{W_iH_i\|U_j-B_i\|^{-\alpha}\right\}\\
 &\hspace{0.75in}> W_lH_l\|U_j-B_l\|^{-\alpha}\bigg],
 \end{align*}
where $\mathcal{E}_{l,j}$ denotes the event that user $U_j$ is not associated with BS $B_l$ as its association region goes to infinity and $\nu(\mathcal{A}_j)$ needs to go to infinity in order to make $\mathbb{P}[\mathcal{E}_{l,j}]$ include all BS association possibilities of user $U_j$. According to Theorem \ref{Thm:RanTransProp} and Remark 1, $\Phi_{\subB}$ can be statistically mapped to another homogeneous PPP $\hat{\Phi}_{\subB}$ of intensity $\hat{\lambda}_{\subB}=\lambda_{\subB}\mathbb{E}[(WH)^{\frac{2}{\alpha}}]$ since $\lambda'=\lambda_{\subB}$ and $T=(WH)^{-1/\alpha}$ in this case, and $\hat{\Phi}_{\subB}(\mathcal{A}_j)$ stands for the number of BSs in $(\mathcal{A}_j\cap\hat{\Phi}_{\subB})$ that is a Poisson random variable with parameter $\hat{\lambda}_{\subB}\nu(\mathcal{A}_j)$. Thus, it follows that                     
 \begin{align*}
\mathbb{P}\left[\mathcal{E}_{l,j}\right]= \lim_{\nu(\mathcal{A}_j)\rightarrow\infty}&\mathbb{P}\bigg[\sup_{\hat{B}_i\in\hat{\Phi}_{\subB}\cap\mathcal{A}_j\setminus \hat{B}_l}\left\{\|\hat{U}_j-\hat{B}_i\|^{-\alpha}\right\}\\
&\hspace{0.2in}>\|\hat{U}_j-\hat{B}_l\|^{-\alpha}\bigg]= \mathbb{P}\left[ \hat{D}_j> \hat{D}^{*}_j\right],
 \end{align*}
 where $\hat{U}_j=(W_lH_l)^{-\frac{1}{\alpha}}U_j$ , $\hat{D}_j$ is the distance from user $\hat{U}_j$ to BS $\hat{B}_l$,  $\hat{D}^{*}_j$ is the distance from  $U_j$ to its nearest BS in $\hat{\Phi}_{\subB}$. Letting $\hat{\mathcal{C}}_l$ be the cell of BS $\hat{B}_l$, the void probability $p_{\emptyset}$ can be found by calculating the following
$$p_{\emptyset}=\mathbb{E}\left\{\mathbb{P}\left[\bigcap_{j=0}^{\Phi_{\subU}(\hat{\mathcal{C}}_l)} \mathcal{E}_{l,j}\bigg| \Phi_{\subU}(\hat{\mathcal{C}}_l)\right]\right\} $$
for all $\hat{\mathcal{C}}_l$'s since $\bigcap_{j=0}^{\Phi_{\subU}(\hat{\mathcal{C}}_l)} \mathcal{E}_{l,j}$ represents the event that BS $\hat{B}_l$ is not associated by any users in its cell $\hat{\mathcal{C}}_l$ and $\Phi_{\subU}(\hat{\mathcal{C}}_l)$ is the random number of users in cell $\hat{\mathcal{C}}_l$. The events $\mathcal{E}_{l,j}$'s could be correlated each other such that the probability $p_{\emptyset}$ is difficult to be explicitly computed. Instead, we find the bounds of  $p_{\emptyset}$ because they can be characterized in closed-form as shown in the following.

First, the lower bound on $p_{\emptyset}$ can be found as follows. By considering a given $\Phi_{\subU}(\hat{\mathcal{C}}_l)$ and any two events $\mathcal{E}_{l,j}$ and $\mathcal{E}_{l,i}$ are highly correlated under $\mathcal{A}_i\cap\mathcal{A}_j\neq\emptyset$ such that $\mathbb{P}[\mathcal{E}_{l,j}\cap \mathcal{E}_{l,i}]\geq\mathbb{P}[\mathcal{E}_{l,i}]\mathbb{P}[\mathcal{E}_{l,j}]$, we have the following
\begin{align*}
\mathbb{P}\left[\bigcap_{j=0}^{\Phi_{\subU}(\hat{\mathcal{C}}_l)} \mathcal{E}_{l,j}\bigg|\Phi_{\subU}(\hat{\mathcal{C}}_l) \right]&= \mathbb{P}\left[\mathcal{E}_{l,0}\cap \mathcal{E}_{l,1}\cap\cdots\cap \mathcal{E}_{l,\Phi_{\subU}(\hat{\mathcal{C}}_l)}\right]\\
&\geq \prod_{j=0}^{\Phi_{\subU}(\hat{\mathcal{C}}_l)}\mathbb{P}[\mathcal{E}_{l,j}].
\end{align*}
Since the number of users in $\hat{\mathcal{C}}_l$ is a Poisson random variable with parameter $\lambda_{\subU}\nu(\hat{\mathcal{C}}_l)$, i.e., $\mathbb{P}[\Phi_{\subU}(\hat{\mathcal{C}}_l)=n]=\frac{(\lambda_{\subU}\nu(\hat{\mathcal{C}}_l))^n}{n!}e^{-\lambda_{\subU}\nu(\hat{\mathcal{C}}_l)}$ for a given $\nu(\hat{\mathcal{C}}_l)$, the lower bound on $p_{\emptyset}$ can be given by
\begin{align}
\mathbb{E}\left\{ \prod_{j=0}^{\Phi_{\subU}(\hat{\mathcal{C}}_l)}\mathbb{P}[\mathcal{E}_{l,j}]\right\}
= \sum_{n=0}^{\infty}\left(\prod_{j=0}^{n}\mathbb{P}[\mathcal{E}_{l,j}]\right) \frac{(\lambda_{\subU}\nu(\hat{\mathcal{C}}_l))^n}{e^{\lambda_{\subU}\nu(\hat{\mathcal{C}}_l)}n!}. \label{Eqn:LowBoundVoidProb02}
\end{align}
According to the Slivnyak theorem \cite{DSWKJM96}, the statistic property evaluated at any point in a homogeneous PPP is the same such that  $\mathbb{P}[\mathcal{E}_{l,j}]$ is the same as $\mathbb{P}[\mathcal{E}_{l,0}]$ evaluated at user $U_0$ (the origin) for all $j$. Accordingly, $\prod_{j=0}^{n}\mathbb{P}[\mathcal{E}_{l,j}]$ in \eqref{Eqn:LowBoundVoidProb02} is equal to $(\mathbb{P}[\mathcal{E}_{l,0}])^n$ and it follows that
\begin{align}
p_{\emptyset} &\geq  \mathbb{E}\left\{e^{-\lambda_{\subU}\nu(\hat{\mathcal{C}}_l)}\sum_{n=0}^{\infty}\left(\mathbb{P}[\mathcal{E}_{l,0}]\right)^n \frac{(\lambda_{\subU}\nu(\hat{\mathcal{C}}_l))^n}{n!}\right\}\nonumber\\
&=\mathbb{E}\left\{ \exp\left[-\lambda_{\subU}\nu(\hat{\mathcal{C}}_l)(1-\mathbb{P}[\mathcal{E}_{l,0}])\right]\right\}.\label{Eqn:LowBoundVoidProb03}
\end{align}
and $\mathbb{P}\left[\mathcal{E}_{l,0}\right]=\mathbb{P}\left[ \hat{D}_0> \hat{D}^*_0\right]
$ in which $\hat{D}^*_0$ is the distance from the origin to the nearest BS in $\hat{\Phi}_{\subB}$. The pdf of $\hat{D}_0^*$ is $f_{\hat{D}_0^*}(x)=2\pi\zeta\lambda_{\subB}x\exp(-\pi\zeta\lambda_{\subB}x^2)$ \cite{FBBBL10}, whereby $\mathbb{P}[\mathcal{E}_{l,0}]$ is simplified as
 \begin{align}
 \mathbb{P}[\mathcal{E}_{l,0}] =1- \mathbb{E}\left[\exp\left(-\pi\zeta\lambda_{\subB}(\hat{D}_0)^2\right)\right].
 \end{align}   
Since all users in $\hat{\mathcal{C}}_l$ are uniformly distributed, we have $f_{\hat{D}_0}(r)=\frac{2\pi r\zeta}{\nu(\hat{\mathcal{C}}_l)}$ and thus we can calculate $\mathbb{P}[\mathcal{E}_{l,0}]$ in closed-form as
\begin{align*}
\mathbb{P}[\mathcal{E}_{l,0}] &= 1-\int_{0}^{\sqrt{\frac{\nu(\hat{\mathcal{C}}_l)}{\pi}}} \exp\left(-\pi\zeta\lambda_{\subB}r^2\right)\frac{2\pi r\zeta }{\nu(\hat{\mathcal{C}}_l)}\dif r\\
&= 1-\frac{1-\exp(-\zeta\lambda_{\subB}\nu(\hat{\mathcal{C}}_l))}{\lambda_{\subB}\nu(\hat{\mathcal{C}}_l)}.
\end{align*}
Clearly, $\frac{1}{\lambda_{\subB}\nu(\hat{\mathcal{C}}_l)} \geq 1-\mathbb{P}[\mathcal{E}_{l,0}]$ and substituting this into \eqref{Eqn:LowBoundVoidProb03} results in the lower bound in  \eqref{Eqn:BoundVoidProb}.

Next, the upper bound on $p_{\emptyset}$ can be obtained as follows. The upper bound on $\mathbb{P}[\mathcal{E}_{l,j}]$ is given by
\begin{align*}
 \mathbb{P}[\mathcal{E}_{l,j}]\leq &\lim_{\nu(\mathcal{A}_j)\rightarrow\infty} \mathbb{P}\bigg[\sup_{B_i\in\Phi_{\subB}\cap\mathcal{A}_j\setminus B_l}W_iH_i\|U_j-B_i\|^{-\alpha}\\
  &\hspace{0.75in}>W_lH_l\inf_{U_j\in\mathcal{C}_l}\{\|U_j-B_l\|^{-\alpha}\}\bigg],
\end{align*}
which is acquired by considering the farthest user of BS $B_l$ in cell $\mathcal{C}_l$.  According to Slivnyak's theorem, for a given  $\Phi_{\subU}(\mathcal{C}_l)$  we can have
$$\mathbb{P}\left[\bigcap_{j=0}^{\Phi_{\subU}(\mathcal{C}_l)} \mathcal{E}_{l,j}\bigg|\Phi_{\subU}(\mathcal{C}_l) \right]\leq \left(\mathbb{P}\left[\hat{R}_l \geq \hat{D}^{*}_0\right]\right)^{\Phi_{\subU}(\mathcal{C}_l)},$$
where $\hat{R}_l$ denotes the distance from BS $\hat{B}_l$ to its farthest user in $\hat{\mathcal{C}}_l$. Letting $\hat{\Phi}_{\subU}(\hat{\mathcal{C}}_l)\defn\{\hat{U}_j\in\hat{\mathcal{C}}_l: (W_jH_j)^{-\frac{1}{\alpha}}U_j, U_j\in\mathcal{C}_l\}$ and $R_{0,j}$ be the distance from BS $B_l$ to the $j$th nearest user in $\mathcal{C}_l$,  we have
\begin{align*}
&\mathbb{P}\left[\hat{R}_l \geq \hat{D}^*_0\right]=\mathbb{P}\left[(\hat{R}_l)^2 \geq (\hat{D}^*_0)^2\right]\\
&\stackrel{(a)}{=}\mathbb{P}\left[(W_lH_l)^{-\frac{2}{\alpha}}\sum^{\hat{\Phi}_{\subU}(\hat{\mathcal{C}}_l)}_{j=1}R_{0,j}^2 \geq (\hat{D}^*_0)^2\right]\\
&\leq \mathbb{E}\left\{\prod_{j=1}^{\hat{\Phi}_{\subU}(\hat{\mathcal{C}}_l)}\mathbb{P}\left[(W_lH_l)^{-\frac{2}{\alpha}}R_{0,j}^2 \geq (\hat{D}^*_0)^2\right]\right\}\\
&= \left(1-\exp\left(-\pi\hat{\lambda}_{\subB} \mathbb{E}\left[(W_lH_l)^{-\frac{2}{\alpha}}\right]R_0^2\right)\right)^{\mathbb{E}[\hat{\Phi}_{\subU}(\hat{\mathcal{C}}_l)]}\\
&\stackrel{(b)}{\leq}  \left(1-\frac{\lambda_{\subU}}{\lambda_{\subU}+\mathbb{E}[(WH)^{-\frac{2}{\alpha}}]\hat{\lambda}_{\subB}}\right)^{\Phi_{\subU}(\mathcal{C}_l)/\zeta}
\end{align*}
\begin{align*}
&=\left(1+\frac{\lambda_{\subU}}{\mathbb{E}[(WH)^{-\frac{2}{\alpha}}]\hat{\lambda}_{\subB}}\right)^{-\Phi_{\subU}(\mathcal{C}_l)/\zeta}\\
 &=\left(1+\frac{\lambda_{\subU}}{\zeta\lambda_{\subB}}\right)^{-\Phi_{\subU}(\mathcal{C}_l)/\zeta}\stackrel{(c)}{\leq}  \left(1+\frac{\lambda_{\subU}}{\zeta\lambda_{\subB}}\right)^{-\zeta/\Phi_{\subU}(\mathcal{C}_l)},
\end{align*}
where (a) follows from that fact that $\{(\hat{R}_l)^2, l=1,2,\cdots\}$ is an one-dimensional Poisson process of intensity $\lambda_{\subU}\mathbb{E}[(WH)^{\frac{2}{\alpha}}]$ and thus $(\hat{R}_l)^2$ is equal to the sum of $\hat{\Phi}_{\subU}(\hat{\mathcal{C}}_l)$ i.i.d. $(WH)^{-\frac{2}{\alpha}}R^2_0$ \cite{MH05}, $(b)$ is due to first applying Jensen's inequality on the term $(WH)^{-\frac{2}{\alpha}}$ and then calculating the expectation regarding $R_0^2$, and $(c)$ is obtained from $(1+x)^n\leq (1+x)^{1/n}$ for $x\geq 0$ and $n\geq 1$ and $\mathbb{E}[\hat{\Phi}_{\subU}(\hat{\mathcal{C}}_l)]=\Phi_{\subU}(\mathcal{C}_l)/\zeta\geq 1$. Therefore, we have
$$\mathbb{P}\left[\bigcap_{j=0}^{\Phi_{\subU}(\mathcal{C}_l)} \mathcal{E}_j\bigg|\Phi_{\subU}(\mathcal{C}_l) \right]\leq\left(1+\frac{\lambda_{\subU}}{\zeta\lambda_{\subB}}\right)^{-\zeta}$$
for any given $\Phi_{\subU}(\mathcal{C}_l)$ and the upper bound in \eqref{Eqn:BoundVoidProb} follows. This completes the proof. 

\subsection{Proof of Proposition \ref{Prop:AvgCellThroughput}}\label{App:ProofAvgCellThroughput}
According to \eqref{Eqn:EquAssoBSofGCA}, we can have the following CDF of $\|B^*_0\|$
\begin{align*}
\mathbb{P}[\|B^{*}_0\|\leq x]&=\mathbb{P}\left[\|\hat{B}_0\|\leq x(W_0H_0)^{-\frac{1}{\alpha}}\right]\\
&=1-\mathbb{E}\left[\exp\left(-\pi\lambda_{\subB}G^2x^2\right)\right],
\end{align*}
where $\hat{B}_0$ is the nearest BS in $\hat{\Phi}_{\subB}$ to the origin. Hence, for a given $W_0H_0$, $\|B^*_0\|$ can be viewed as the distance from the nearest BS in $\Phi_{\subB}$ to the origin that is scaled by $1/G$. That means $I_0\|B^*_0\|^{\alpha}=(G^{-\alpha}I_0)(G^{\alpha}\|B^*_0\|^{\alpha})\stackrel{d}{=}G^{-\alpha}I_0\|B_0\|^{\alpha}$ if $B_0$ is the nearest BS in $\Phi_{\subB}$ to the origin. For a given $\|B_0\|=u$ and $G$, the Laplace functional of $G^{-\alpha}I_0u^{\alpha}$ can be found by
\begin{align*}
&\mathcal{L}_{u^{\alpha}I_0G^{-\alpha}}(s|u)=\mathbb{E}\left[e^{-s (u/G)^{\alpha}I_0}\right]\\
&\stackrel{(a)}{=}\mathbb{E}\left[e^{-sG^{-\alpha}\sum_{B_{i-1}\in\Phi_{\subB}}V_iH_i [u^2/(\|B_{i-1}\|^2+u^2)]^{\frac{\alpha}{2}}}\right]\\
&\stackrel{(b)}{=}e^{-\mathbb{E}\left[\int_{0}^{\infty}\left(1-e^{-sG^{-\alpha}VH\left(1+\frac{r^2}{u^2}\right)^{-\frac{\alpha}{2}}}\right)2\pi\lambda_{\subB}r\dif r\right]}\\
&\stackrel{(c)}{=} e^{-\pi\lambda_{\subB} u^2(1-p_{\emptyset})\mathbb{E}\left[\int_{1}^{\infty}\left(1-e^{-sG^{-\alpha}Hx^{-\frac{\alpha}{2}}}\right) \dif x\right]}
\end{align*}
where $(a)$ is due to the fact that $\{\|B_i\|^2, i=1,2,\ldots\}$ are an one-dimensional Poisson process of intensity $\pi\lambda_{\subB}$, $(b)$ follows from the probability generation function of a homogeneous PPP \cite{DSWKJM96} and Theorem \ref{Thm:RanTransProp} and $(c)$ is due to assuming the associated BSs are a thinning PPP of intensity $\lambda_{\subB}(1-p_{\emptyset})$. By assuming $Y$ is an exponential random variable with unit mean and variance, we have the following identity:
\begin{align*}
&\int_{0}^{\infty}\left(1-e^{-sG^{-\alpha}Hx^{-\frac{\alpha}{2}}}\right) \dif x =\int_{0}^{\infty} \mathbb{P}\left[Y\leq \frac{sH}{G^{\alpha}x^{\frac{\alpha}{2}}}\right]\dif x\\
&= \int_{0}^{\infty} \mathbb{P}\left\{x\leq \left(\frac{sH}{G^{\alpha}Y}\right)^{\frac{2}{\alpha}}\right\}\dif x=\mathbb{E}_Y\left\{\left(\frac{sH}{G^{\alpha}Y}\right)^{\frac{2}{\alpha}}\right\}
\end{align*}
\begin{align*}
&=(sH)^{\frac{2}{\alpha}}G^{-2}\Gamma\left(1-\frac{2}{\alpha}\right),
\end{align*}
which leads to the following result 
\begin{align}
\mathcal{L}_{u^{\alpha}G^{-\alpha}I_0}(s|u) =e^{-\pi\lambda_{\subB} u^2(1-p_{\emptyset})G^{-2}\ell(s,G) }. \label{Eqn:ProofLapInter}
\end{align}
Then $\mathcal{L}_{\|B^*_0\|^{\alpha}I_0}(s)$ in \eqref{Eqn:LapInter} can be obtained by calculating the following integral
\begin{align*}
\mathcal{L}_{\|B^*_0\|^{\alpha}I_0}(s)&=\mathbb{E}\left[\int_{0}^{\infty}\mathcal{L}_{u^{\alpha}G^{-\alpha}I_0}(s|u) e^{-\pi\lambda_{\subB} u^2} \dif (\pi\lambda_{\subB} u^2)\right]\\
&=\mathbb{E}\left[\frac{G^2}{G^2+(1-p_{\emptyset})\ell(s,G^2)}\right].
\end{align*}
Since $\mathcal{L}_{\|B^*_0\|^{\alpha}I_0}(s)$  is concave for $G^2$, we can use Jensen's inequality to obtain  $\sup\{\mathcal{L}_{\|B^*_0\|^{\alpha}I_0}(s)\}=\frac{1}{1+(1-p_{\emptyset})\ell(s,\zeta)/\zeta}$.

Now we find $\mathcal{T}_{\subC}$. According to \eqref{Eqn:ApporxPdfVoronoiCell}, Proposition \ref{Prop:VoidCellProbGCA} and the assumption that all associated BSs are a homogeneous PPP,  the pdf of the area of associated cell $C^*_0$ can be written as
\begin{equation}\label{Eqn:PdfAssCell}
f_{\nu(C^*_0)} (x) = \frac{(\rho\lambda_{\subB}(1-p_{\emptyset})x)^{\rho}}{\Gamma(\rho)x}e^{-\rho\lambda_{\subB}(1-p_{\emptyset})x}
\end{equation}
and the mean of $1/\nu(C_0^*)$ can be found by
\begin{equation}\label{Eqn:MeanInvSizeAssCell}
\mathbb{E}[1/\nu(C_0^*)] = \int_{0^+}^{\infty} \frac{1}{x}f_{\nu(C_0^*)}(x) \dif x =\frac{\rho \lambda_{\subB}(1-p_{\emptyset})}{\rho-1}.
\end{equation}
since $\rho\geq\frac{7}{2}$. Using this result, the average cell throughput in \eqref{Eqn:AvgCellThroughput} can be expressed as
\begin{align*}
\mathcal{T}_{\subC}&=\mathbb{E}\left[\log_2\left(1+\frac{H_0}{I_0\|B_0^*\|^{\alpha}}\right)\right]\cdot\mathbb{E}\left[\frac{1}{\nu(C_0^*)}\right]\\
&= \frac{\rho\lambda_{\subB}(1-p_{\emptyset})}{(\ln 2)(\rho-1)}\int_{0}^{\infty} \mathbb{P}\left[\frac{H_0}{\|B_0^*\|^{\alpha}I_0}\geq (2^x-1)\right] \dif x \\
&=\frac{\lambda_{\subB}(1-p_{\emptyset})}{(\ln 2)(1-1/\rho)}\int_{0}^{\infty} \mathbb{P}\left[\frac{H_0}{\|B_0^*\|^{\alpha}I_0}\geq y\right] \frac{\dif y}{(y+1)}.
\end{align*}
Moreover, we know
\begin{align}
\mathbb{P}[H>h] &=\frac{1}{\sqrt{2\pi}}\int_{-\infty}^{\infty}e^{-he^{-z}-\frac{(z-\mu_s)^2}{2\sigma^2_s}} \dif z  \nonumber\\
&=\mathbb{E}\left[e^{-he^{-Z}}\right]=\mathcal{L}_{e^{-Z}}(h),
\end{align}
where $Z$ is a normal random variable with mean $\mu_s$ and variance $\sigma^2_s$. Thus, $\mathcal{T}_{\subU}$ can be expressed as
\begin{align*}
\mathcal{T}_{\subC} &=\frac{\rho\lambda_{\subB}(1-p_{\emptyset})}{(\ln 2)(\rho-1)}\mathbb{E} \left[\int_{0}^{\infty}\frac{\exp\left(-yI_0\|B_0^*\|e^{-Z}\right)}{(y+1)} \dif y\right]\\
&=\frac{\lambda_{\subB}(1-p_{\emptyset})}{(\ln 2)(1-1/\rho)}\int_{0}^{\infty}\mathbb{E} \left[\frac{\mathcal{L}_{I_0\|B_0^*\|}(s) }{(s+e^{-Z})}\right]\dif s.
\end{align*}
Using the above result and the definition of $\mathcal{T}_{\subC}$, the average cell throughput can be further written as
\begin{align}
\mathcal{T}_{\subC}=\frac{\lambda_{\subB}(1-p_{\emptyset})}{(\ln 2)(1-1/\rho)}\int_{0}^{\infty}\frac{\mathbb{E} \left[(s+e^{-Z})^{-1}\right]}{1+(1-p_{\emptyset})\ell(s,\zeta)/\zeta} \dif s. \label{Eqn:ProofCellThrou}
\end{align}
Also, for any $s>0$ we have
\begin{align}
\mathbb{E}\left[\frac{1}{s+e^{-Z}}\right]&=\frac{1}{\sqrt{\pi}}\int_{-\infty}^{\infty} \left(s+e^{-(\sqrt{2}\sigma_s x+\mu_s)}\right)^{-1} e^{-x^2} \dif x \nonumber\\
&\stackrel{(d)}{=} \sum_{i=1}^{n}  \frac{\omega_i}{s+e^{-(\sqrt{2}\sigma_s x_i+\mu_s)}}, \label{Eqn:ProofExpInvLogNorm}
\end{align}
where $(d)$ is obtained by using a Gauss-Hermite quadrature integration with a sufficiently large integer $n$. Substituting \eqref{Eqn:ProofExpInvLogNorm} into \eqref{Eqn:ProofCellThrou} results in \eqref{Eqn:AvgCellThroughput}. For the case of no shadowing, letting $Z$ in \eqref{Eqn:ProofCellThrou} be zero gives \eqref{Eqn:AvgCellThroughputNoShadow}. This completes the proof. 

\subsection{Proof of Proposition \ref{Prop:OptGreenCellLoad}}\label{App:OptGreenCellLoad}
The terms in $\mathcal{G}_{\subC}$ that pertain to the cell load can be lumped into the following sole function
$$\mathfrak{V}(v)=\frac{\tilde{p}_{\emptyset}}{(1+\tilde{p}_{\emptyset}\ell(s,\zeta)/\zeta)(\tilde{p}_{\emptyset}(P_{\subON}+\kappa v^{\frac{\alpha}{2}})+(1-\tilde{p}_{\emptyset})P_{\subOFF})},$$
where $v$ represents cell load, $\tilde{p}_{\emptyset}=1-p_{\emptyset}(v)$ and $\kappa=\delta P_{\min}\Gamma(1+\frac{2}{\alpha})/(\pi\lambda_{\subU}\zeta)^{\frac{\alpha}{2}}$ is a constant for a given $\lambda_{\subU}$. Then taking the derivative of $\mathfrak{V}(v)$ w.r.t. $v$ and making it equal to zero yield the following differential equation:
\begin{align}
&\left[\frac{P_{\subOFF}}{\tilde{p}^2_{\emptyset}}-\ell(s,\zeta)(P_{\subON}-P_{\subOFF})\right]\frac{\dif \tilde{p}_{\emptyset}}{\dif v}=\frac{\kappa\ell(s,\zeta) v^{\frac{\alpha}{2}}}{\zeta}\frac{\dif \tilde{p}_{\emptyset}}{\dif v}\nonumber\\
&+\frac{\alpha\kappa}{2}v^{\frac{\alpha}{2}-1}\left(1+\frac{\ell(s,\zeta)\tilde{p}_{\emptyset}}{\zeta}\right)
=\kappa \frac{\dif}{\dif v}\left[v^{\frac{\alpha}{2}}(1+\tilde{p}_{\emptyset}\ell(s,\zeta)/\zeta)\right].\label{Eqn:DiffEqnGreenCellLoad}
\end{align}
Integrating both sides of the equation above and applying the appropriate boundary conditions of $\tilde{p}_{\emptyset}$ under the constraint that $\tilde{p}_{\emptyset}$ is not a function of $\ell(s,\alpha)$ lead to\footnote{Please refer to the proof of Proposition \ref{Prop:OptCellLoadUserThr} for the details in how to determine the exact solution form of this kind of differential equation.} 
\begin{align*}
&\tilde{p}^{\frac{1}{\beta}-1}_{\emptyset}(P_{\subON}-P_{\subOFF})=\kappa v^{\frac{\alpha}{2}}+(P_{\subON}-P_{\subOFF})\\
\Rightarrow& v^{\frac{\alpha}{2}}=\frac{(P_{\subON}-P_{\subOFF})}{\kappa}\left(\tilde{p}^{\frac{1}{\beta}-1}_{\emptyset}-1\right).
\end{align*}
Since $\tilde{p}_{\emptyset}=1-(1+v/\rho)^{-\rho}$, $\mathds{L}_{\subC}(v)$ in \eqref{Eqn:OptGreenCellLoad} has a fixed point at $v^*$. 

Now wee need to show the uniqueness of the solution of the differential equation in \eqref{Eqn:DiffEqnGreenCellLoad}, i.e., the fixed point of $\mathds{L}_{\subC}(v)$ is unique. First, we rewrite \eqref{Eqn:DiffEqnGreenCellLoad} as
$$\frac{\dif\tilde{p}_{\emptyset}}{\dif\left[v^{\frac{\alpha}{2}}(1+\tilde{p}_{\emptyset}\ell(s,\zeta)/\zeta)\right] }=\frac{\kappa\tilde{p}^2_{\emptyset}}{P_{\subOFF}-(P_{\subON}-P_{\subOFF})\tilde{p}^2_{\emptyset}\ell(s,\zeta)/\zeta}.$$
Assume both $\psi_1(v)$ and $\psi_2(v)$ are two arbitrary solutions of \eqref{Eqn:DiffEqnGreenCellLoad} and let $\tilde{\psi}(v)=\psi_1-\psi_2$. Since $0\leq \tilde{p}_{\emptyset}\leq 1$, the following inequality
$$|\tilde{\psi}|\leq \int_{0}^{v^{\frac{\alpha}{2}}(1+\ell(s,\zeta))} \frac{\kappa |\tilde{\psi}|\dif \tau}{|P_{\subOFF}-\ell(s,\alpha)(P_{\subON}-P_{\subOFF})|}$$
always holds. By the Gronwall-Bellman lemma \cite{WJR96}, we know $|\tilde{\psi}|\leq 0$ and thus $\tilde{\psi}=0$. Consequently, the solution is unique since any two arbitrary solutions $\psi_1$ and $\psi_2$ are equal.   

\bibliographystyle{ieeetran}
\bibliography{IEEEabrv,Ref_CellAssOptDep}

\begin{thebibliography}{10}
\providecommand{\url}[1]{#1}
\csname url@samestyle\endcsname
\providecommand{\newblock}{\relax}
\providecommand{\bibinfo}[2]{#2}
\providecommand{\BIBentrySTDinterwordspacing}{\spaceskip=0pt\relax}
\providecommand{\BIBentryALTinterwordstretchfactor}{4}
\providecommand{\BIBentryALTinterwordspacing}{\spaceskip=\fontdimen2\font plus
\BIBentryALTinterwordstretchfactor\fontdimen3\font minus
  \fontdimen4\font\relax}
\providecommand{\BIBforeignlanguage}[2]{{%
\expandafter\ifx\csname l@#1\endcsname\relax
\typeout{** WARNING: IEEEtran.bst: No hyphenation pattern has been}%
\typeout{** loaded for the language `#1'. Using the pattern for}%
\typeout{** the default language instead.}%
\else
\language=\csname l@#1\endcsname
\fi
#2}}
\providecommand{\BIBdecl}{\relax}
\BIBdecl

\bibitem{NBJLDM14}
N.~Bhushan, J.~Li, D.~Malladi \emph{et~al.}, ``Network densification: the
  dominant theme for wireless evolution into 5{G},'' \emph{{IEEE} Commun.
  Mag.}, vol.~52, no.~2, pp. 82--89, Feb. 2014.

\bibitem{IHBSSSS13}
I.~Hwang, B.~Song, and S.~Soliman, ``A holistic view on hyper-dense
  heterogeneous and small cell networks,'' \emph{{IEEE} Commun. Mag.}, vol.~51,
  no.~6, pp. 20--27, Jun. 2013.

\bibitem{AFGFJMGB11}
A.~Fehske, G.~Fettweis, J.~Malmodin, and G.~Biczok, ``The global footprint of
  mobile communications: The ecological and economic perspective,''
  \emph{{IEEE} Commun. Mag.}, vol.~49, no.~8, pp. 55--62, Aug. 2011.

\bibitem{IAFBLH11}
I.~Ashraf, F.~Boccardi, and L.~Ho, ``{SLEEP} mode techniques for small cell
  deployments,'' \emph{{IEEE} Commun. Mag.}, vol.~49, no.~8, pp. 72--79, Aug.
  2011.

\bibitem{JGAFBRKG11}
J.~G. Andrews, F.~Baccelli, and R.~K. Ganti, ``A tractable approach to coverage
  and rate in cellular networks,'' \emph{{IEEE} Trans. Commun.}, vol.~59,
  no.~11, pp. 3122--3134, Nov. 2011.

\bibitem{HSDRKGFBJGA12}
H.~S. Dhillon, R.~K. Ganti, F.~Baccelli, and J.~G. Andrews, ``Modeling and
  analysis of {K}-tier downlink heterogeneous cellular networks,'' \emph{{IEEE}
  J. Sel. Areas Commun.}, vol.~30, no.~3, pp. 550 -- 560, Apr. 2012.

\bibitem{HSJYJSXPJGA12}
H.-S. Jo, Y.~J. Sang, X.~Ping, and J.~G. Andrews, ``Heterogeneous cellular
  networks with flexible cell association: A comprehensive downlink {SINR}
  analysis,'' \emph{{IEEE} Trans. Wireless Commun.}, vol.~11, no.~10, pp.
  3484--3495, Oct. 2012.

\bibitem{PXCHLJGA13}
P.~Xia, C.-H. Liu, and J.~G. Andrews, ``Downlink coordinated multi-point with
  overhead modeling in heterogeneous cellular networks,'' \emph{{IEEE} Trans.
  Wireless Commun.}, vol.~12, no.~8, pp. 4025--4037, Jun. 2013.

\bibitem{MDRAGGEC13}
M.~D. Renzo, A.~Guidotti, and G.~E. Corazza, ``Average rate of downlink
  heterogeneous cellular networks over generalized fading channels: A
  stochastic geometry approach,'' \emph{{IEEE} Trans. Commun.}, vol.~61, no.~7,
  pp. 3050--3071, Jul. 2013.

\bibitem{SSJGA14}
S.~Singh and J.~G. Andrews, ``Joint resource partitioning and offloading in
  heterogeneous cellular networks,'' \emph{{IEEE} Trans. Wireless Commun.},
  vol.~13, no.~2, pp. 888--901, Feb. 2014.

\bibitem{SLKH12}
S.~Lee and K.~Huang, ``Coverage and economy of cellular networks with many base
  stations,'' \emph{{IEEE} Commun. Lett.}, vol.~16, no.~7, pp. 1038--1040, Jul.
  2012.

\bibitem{CLJZKBL14}
C.~Li, J.~Zhang, and K.~B. Letaief, ``Throughput and energy efficiency analysis
  of small cell networks with multi-antenna base stations,'' \emph{{IEEE}
  Trans. Wireless Commun.}, vol.~13, no.~5, pp. 2505--2517, May 2014.

\bibitem{CTPCHLLCW15}
C.-T. Peng, C.-H. Liu, and L.-C. Wang, ``Optimal base station deployment for
  small cell networks with energy-efficient power control,'' in \emph{IEEE
  International Conference on Communications}, Jun. 2015.

\bibitem{HSMSMWTQSQ14}
H.~Sun, M.~Sheng, M.~Wildemeersch, and T.~Q.~S. Quek, ``Coverage analysis for
  two-tier dynamic tdd heterogeneous networks,'' in \emph{IEEE Global
  Communication Conference}, Dec. 2014.

\bibitem{HSMWMSTQSQ15}
H.~Sun, M.~Wildemeersch, M.~Sheng, and T.~Q.~S. Quek, ``D2d enhanced
  heterogeneous cellular networks with dynamic tdd,'' \emph{To appear in IEEE
  Trans. on Wireless Communications}.

\bibitem{TQSQWCCMK11}
T.~Q.~S. Quek, W.~C. Cheung, and M.~Kountouris, ``Energy efficiency analysis of
  two-tier heterogeneous networks,'' in \emph{the 11th European Wireless
  Conference}, Apr. 2011, pp. 1--5.

\bibitem{DCSZZN13}
D.~Cao, S.~Zhou, and Z.~Niu, ``Optimal combination of base station densities
  for energy-efficient two-tier heterogeneous cellular networks,'' \emph{{IEEE}
  Trans. Wireless Commun.}, vol.~12, no.~9, pp. 4350--4362, Sep. 2013.

\bibitem{MDAGGEC13}
M.~D. Renzo, A.~Guidotti, and G.~E. Corazza, ``Average rate of downlink
  heterogeneous cellular networks over generalized fading channels: A
  stochastic geometry approach,'' \emph{{IEEE} Trans. Commun.}, vol.~61, no.~7,
  pp. 3050--3071, Jul. 2013.

\bibitem{SRCWC13}
S.-R. Cho and W.~Choi, ``Energy-efficient repulsive cell activation for
  heterogeneous cellular networks,'' \emph{{IEEE} J. Sel. Areas Commun.},
  vol.~31, no.~5, pp. 870--882, May 2013.

\bibitem{YSSTQSQMKHS13}
Y.~S. Soh, T.~Q.~S. Quek, M.~Kountouris, and H.~Shin, ``Energy efficient
  heterogeneous cellular networks,'' \emph{{IEEE} J. Sel. Areas Commun.},
  vol.~31, no.~5, pp. 840--850, May 2013.

\bibitem{STUBER01}
G.~L. St\"{u}ber, \emph{Principles of Mobile Communication}, 2nd~ed.\hskip 1em
  plus 0.5em minus 0.4em\relax Springer, 2011.

\bibitem{DSWKJM96}
D.~Stoyan, W.~Kendall, and J.~Mecke, \emph{Stochastic Geometry and Its
  Applications}, 2nd~ed.\hskip 1em plus 0.5em minus 0.4em\relax New York: John
  Wiley and Sons, Inc., 1996.

\bibitem{CHLBRSC15}
C.-H. Liu, B.~Rong, and S.~Cui, ``Optimal discrete power control in
  {P}oisson-clustered ad hoc networks,'' \emph{{IEEE} Trans. Wireless Commun.},
  vol.~14, no.~1, pp. 138--151, Jan. 2015.

\bibitem{GAVGCD11}
G.~Auer, V.~Giannini, C.~Desset \emph{et~al.}, ``How much energy is needed to
  run a wireless network?'' \emph{{IEEE} Commun. Mag.}, vol.~18, no.~5, pp.
  40--49, Oct. 2011.

\bibitem{JSFNZ07}
J.-S. Ferenc and Z.~N{\'e}da, ``On the size distribution of {P}oisson {V}oronoi
  cells,'' \emph{Physica A: Statistical Mechanics and its Applications}, vol.
  385, no.~2, pp. 518--526, 2007.

\bibitem{MAIAS72}
M.~Abramowitz and I.~A. Stegun, \emph{Handbook of Mathematical Functions: with
  Formulas, Graphs, and Mathematical Tables}, 9th~ed.\hskip 1em plus 0.5em
  minus 0.4em\relax Dover Publications, 1972.

\bibitem{CHLLCW15}
C.-H. Liu and L.-C. Wang, ``Random cell association and void probability in
  poisson-distributed cellular networks,'' in \emph{IEEE International
  Conference on Communications}, Jun. 2015.

\bibitem{3GPP36.814.10}
{3GPP TR 36.814 v9}, ``{Further advancements for E-UTRA physical layer
  aspects},'' Mar. 2010.

\bibitem{3GPP36.872.13}
{3GPP TR 36.872 v12}, ``{Small cell enhancements for E-UTRA and E-UTRAN -
  Physical layer aspects},'' Sep. 2013.

\bibitem{FBBBL10}
F.~Baccelli and B.~B{\l}aszczyszyn, ``Stochastic geometry and wireless
  networks: Volume {I T}heory,'' \emph{Foundations and Trends in Networking},
  vol.~3, no. 3-4, pp. 249--449, 2010.

\bibitem{MH05}
M.~Haenggi, ``On distances in uniformly random networks,'' \emph{{IEEE} Trans.
  Inf. Theory}, vol.~51, no.~10, pp. 3584 --3586, October 2005.

\bibitem{WJR96}
W.~J. Rugh, \emph{Linear System Theory}.\hskip 1em plus 0.5em minus 0.4em\relax
  Upper Saddle River, NJ 07458: Prentice Hall, 1996.

\end{thebibliography}

\end{document}